\documentclass[sigconf]{acmart}
\renewcommand\footnotetextcopyrightpermission[1]{}
\usepackage{times}
\usepackage{adjustbox}
\usepackage{graphics}
\usepackage{graphicx}
\usepackage{subfig}
\usepackage{natbib}
\usepackage{algorithm}
\usepackage{algorithmic}
\usepackage{todonotes}
\usepackage{amsmath}
\usepackage{amsfonts}
\usepackage{amssymb}
\usepackage{pifont}

\newcommand{\RR}{\mathbb{R}}

\usepackage{graphicx}
\usepackage{balance}

\title[FT-Caffe]{What does fault tolerant Deep Learning need from MPI?}

\begin{document}
\setcopyright{none}

\author{Vinay Amatya}
\affiliation{%
\institution{Pacific Northwest National Lab}
\streetaddress{902 Battelle Blvd}
\city{Richland}
\state{WA}
\postcode{99352}
}
\email{vinay.amatya@pnnl.gov}

\author{Abhinav Vishnu}
\affiliation{%
\institution{Pacific Northwest National Lab}
\streetaddress{902 Battelle Blvd}
\city{Richland}
\state{WA}
\postcode{99352}
}
\email{abhinav.vishnu@pnnl.gov}

\author{Charles Siegel}
\affiliation{%
\institution{Pacific Northwest National Lab}
\streetaddress{902 Battelle Blvd}
\city{Richland}
\state{WA}
\postcode{99352}
}
\email{charles.siegel@pnnl.gov}
\author{Jeff Daily}
\affiliation{%
\institution{Pacific Northwest National Lab}
\streetaddress{902 Battelle Blvd}
\city{Richland}
\state{WA}
\postcode{99352}
}
\email{jeff.daily@pnnl.gov}

\begin{abstract}
Deep Learning (DL) algorithms have become the {\em de facto} Machine Learning
(ML) algorithm for large scale data analysis.  DL algorithms are
computationally expensive -- even distributed DL implementations which use MPI
require days of training (model learning) time on commonly studied datasets.
Long running DL applications become susceptible to faults -- requiring
development of a fault tolerant system infrastructure, in addition to fault
tolerant DL algorithms. This raises an important question: {\em What is needed
from MPI for designing fault tolerant DL implementations?}  In this paper, we
address this problem for permanent faults. We motivate the need for a fault
tolerant MPI specification by an in-depth consideration of recent innovations
in DL algorithms and their properties, which drive the need for specific fault
tolerance features. We present an in-depth discussion on the suitability of
different parallelism types (model, data and hybrid); a need (or lack thereof)
for check-pointing of any critical data structures; and most importantly,
consideration for several fault tolerance proposals (user-level fault
mitigation (ULFM), Reinit) in MPI and their applicability to fault tolerant DL
implementations. We leverage a distributed memory implementation of Caffe,
currently available under the Machine Learning Toolkit for Extreme Scale
(MaTEx). We implement our approaches by extending MaTEx-Caffe for using
ULFM-based implementation.  Our evaluation using the ImageNet dataset and
AlexNet, and GoogLeNet neural network topologies demonstrates the effectiveness of the proposed
fault tolerant DL implementation using OpenMPI based ULFM.

\end{abstract}

\maketitle

\section{Introduction}
\label{sec:intro}
Deep Learning (DL) algorithms are a class of Machine Learning and Data Mining
algorithms, which conduct model learning by emulating the computational
structure of a mammalian brain. A deep neural network (DNN) -- which stores the
model of a DL algorithm -- contains several {\em layers} of {\em neurons}
inter-connected with {\em synapses}. By using deep layers, DL algorithms are
able to conduct transformations on highly non-linear data, which is commonplace
in many scientific datasets.  DL algorithms have shown amazing results in many
Computer Vision tasks~\cite{NIPS2012_4824,43022} and science domains such as
High Energy Physics~\cite{Baldi:2014kfa}, Climate
Modeling~\cite{liu2016application} and Computational Chemistry~\cite{goh2017deep}. DL implementations such as
TensorFlow~\cite{tensorflow2015-whitepaper}, Caffe~\cite{jia2014caffe},
Theano~\cite{bergstra+al:2010-scipy, Bastien-Theano-2012}, and
Torch~\cite{Collobert02torch:a} have become available. These implementations
are primarily geared towards compute nodes that may contain a multi-core
architecture (such as Intel Xeon/KNC/KNL) and/or many-core architectures (GPUs)
as commonplace in Leadership Class Facilities (LCFs).

DL algorithms can be applied to a variety of input representations. The tabular
input representations typically leverage Multi-layer Perceptrons (MLPs). The
images, videos and speech tend to leverage the Convolutional Neural Networks
(CNNs) and Recurrent Neural Networks (RNNs). The CNNs and RNNs are
computationally expensive and typically require significant time for training
even on relatively modest data set sizes with modest number of hidden layers.
The problem is further exacerbated by the increasing number of layers (such as
recently proposed Residual Networks have up to 1000 layers) and ever-increasing
volume of data produced by simulations, experiments and hand-held devices.  An
important solution to these problems is the design and implementation of DL
algorithms that are capable of execution on distributed memory systems.
Table~\ref{table:newcomparisons} shows a table of prominent distributed DL
implementations.

\begin{table}[!h]
	\centering
		\begin{tabular}{|c|c|c|}
		\hline
		Name  & HPC Ready  & Fault Tolerance \\
		\hline
		FireCaffe~\cite{firecaffe} & \checkmark & \ding{55}  \\
		S-Caffe~\cite{scaffe} & \checkmark & \ding{55}  \\
		MaTEx~\cite{matex} & \checkmark & \ding{55}  \\
		Poseidon~\cite{poseidon} & \ding{55} & \ding{55}  \\
		Petuum~\cite{petuum} & \ding{55} & \ding{55}  \\
		GeePS~\cite{geeps} &  \ding{55} & ?  \\
		ProjectAdam~\cite{projectadam} & \ding{55} & \ding{55}  \\
		TensorFlow~\cite{tensorflow2015-whitepaper} &  \ding{55} & \checkmark  \\
		MXNET~\cite{mxnet} &  \ding{55} & \checkmark  \\
		CaffeonSpark~\cite{caffeonspark} & \ding{55} & \checkmark  \\
		SparkNet~\cite{sparknet} &  \ding{55} & \checkmark  \\
		DogWild~\cite{sparknet}  & \ding{55} & ?  \\
		CNTK~\cite{cntk} &  \checkmark & \ding{55}  \\
		Parle~\cite{parle} & \ding{55} & \ding{55}  \\
		PaddlePaddle~\cite{mxnet} &  \checkmark & \ding{55}  \\
		Caffe2~\cite{goyal:arxiv17} & \checkmark & \ding{55}  \\
		\hline
		Proposed FT-Caffe & \checkmark & \checkmark  \\
		\hline
		\hline
	\end{tabular}\\
	\caption{\small{Comparison of the proposed fault tolerant DL with prominent published ppproaches.
	HPC ready implementations leverage the HPC interconnects natively such as either using MPI or native interfaces.
	}}
	\label{table:newcomparisons}
\end{table}

An important artifact of the large scale systems is the increased frequency of
faults, which are commonplace in large scale systems~\cite{Schroeder:4775906}.
Distributed DL implementations such as distributed TensorFlow, distributed
memory implementations of Caffe and even recently proposed Caffe2
~\cite{Caffe2} are primarily geared towards performance.  As shown in
Table~\ref{table:newcomparisons}, we observe that state of the art HPC ready DL
implementations are not fault tolerant. On the other hand, automatic fault
tolerance is provided by MapReduce instantiations such as Hadoop, and Spark.
However, the implementations are not HPC ready.  At the same time, DL
implementations are known to take days even on modest size datasets,
significantly increasing the probability of observing a fault during the training
phase. This raises two important questions: {\em 1) What are the elements of
fault tolerant DL algorithms? and  2) What is needed from MPI for
implementing these fault tolerant DL algorithms?}

\subsection{Contributions}
In this paper, we address these questions and make the following contributions:
\begin{itemize}
\item We present the case for several types of parallelism (model, data and hybrid)
	as motivated
from common use-cases and DNN topologies. We use this
discussion to derive the suitability of fault tolerance
proposals in MPI.
\item We consider several design choices for implementing fault
tolerant DL implementations. Specifically, we consider
checkpoint-restart, Reinit (when a fault occurs, re-initialize
the MPI automatically) and user-level fault mitigation.
\item We consider several approaches for recovery from faults. We
primarily rely on ``continued execution'' -- where the DL
implementation continues to execute by using the remaining set of
compute nodes.
\item We implement our design using MaTEx-Caffe and leverage the ULFM
implementation available with OpenMPI. We provide an evaluation
of fault tolerant MaTEx-Caffe using the ImageNet-1K dataset and
widely studied neural network topologies such as AlexNet and GoogLeNet.
\end{itemize}

Our evaluation on a 16 node Intel Haswell system connected with InfiniBand
indicates that the proposed fault tolerant MaTEx-Caffe is able to scale well
and continues execution in the presence of actual permanent node faults.  It
incurs no observable overhead in the absence of faults, and provides expected
performance after recovering from faults, since the overall number of compute
nodes are reduced.  We also observe that both Reinit~\cite{Reinit} and
ULFM~\cite{Bland2013} proposals are suitable for addressing permanent node
faults for DL algorithms. However, ULFM is simple -- and versatile enough --
since it obviates the need for any checkpoint/restart, re-reading of the entire
dataset and allows continued execution in the presence of permanent node
faults.

The rest of the paper is organized as follows: In section~\ref{sec:background},
we present the background of the proposed research. We present elements of
scalable DL algorithms in section~\ref{sec:motivation} and a solution space
for fault tolerant MaTEx-Caffe in section~\ref{sec:design}. In
section~\ref{sec:existing_proposals} we
present existing features and proposals for fault tolerance in MPI and
provide implementation
details in section~\ref{sec:implDetail}. We discuss experimental results in
section~\ref{sec:exp}, related work in section~\ref{sec:related} followed by
conclusions in section~\ref{sec:conclusions}.

\section{Background}
\label{sec:background}

In this section, we provide a brief overview of the proposed research.
Specifically, we focus on deep neural networks (DNNs).

\subsection{Deep Neural Networks}

\begin{table}[!htbp]
  \centering
	\begin{tabular}{|c|c|c|c|c|c|c|c|c|c|}
    \hline
		Symbol  & Meaning\\
		\hline
		$W^{(\ell)}$		&weights of layer $\ell$\\
		$b^{(\ell)}$		&biases of layer $\ell$\\
		$z^{(\ell+1)}$	&$W^{(\ell)}a^{(\ell)}+b^{(\ell)}$\\
		$a^{(\ell)}$		&$\sigma(z^{(\ell)})$\\
		$\sigma$				&$\mathrm{ReLU}(x)=\max(0, x)$\\
		$n_\ell$				&number of layers\\
		\hline
	\end{tabular}\\
	\caption{Symbols for Backpropagation.}
	\label{table:notation}
\end{table}

A dataset is a collection of samples.  Samples are often images, speech, text
or raw vectors of numbers.  ML algorithms typically split a dataset into a {\em
training set}, used for learning the details of a model; a {\em validation
set}, used to prevent overfitting and to tune hyper-parameters; and a {\em
testing set}, used for the accuracy calculation after the final model is
trained.  Deep neural networks (DNNs) are a class of ML algorithm that learn
nonlinear functions by emulating the computational structure of a mammalian
brain.  It consists of
simple computational units called {\em neurons} which are connected with {\em
synapses}.

The values of the synapses, called {\em weights} are learned through the {\em
back-propagation} algorithm.  It iteratively updates the weights of the DNN to
find a local minimum of an objective/ cost function.  With this algorithm, each sample is
an input to the {\em feed-forward} step.  The output is a {\em predicted value}
which is compared to a {\em label}.  The difference between the label and predicted value  is used to
calculate the {\em gradients} which are applied to update the weights.  This
difference is called the {\em cost} and the objective of training is to
minimize this value on the training set while ensuring that the value on the
validation set decreases as well.

Back-propagation is a special case of {\em gradient descent}.  Any gradient
descent variant uses the update rule
\begin{eqnarray}
	\label{graddesc}
	\mathbf{w}'&=&\mathbf{w}+\lambda \nabla_{\mathbf{w}}C\\
	\mathbf{b}'&=&\mathbf{b}+\lambda \nabla_{\mathbf{b}}C.
\end{eqnarray}
where $\mathbf{w}$ are the weights, $\mathbf{b}$ the biases, $\lambda$ the
learning rate, and $C$ is a cost function to be optimized.

We use the notation of Table~\ref{table:notation} and describe back-propagation in Algorithm~\ref{alg:bp}.

\begin{algorithm}
	\caption{Back Propagation~\cite{marsland2015machine}}
	\begin{algorithmic}[1]
		\label{alg:bp}
		\STATE \textbf{input:} Data $X\in \RR^{n\times p}$ and labels $Y\in \RR^{n\times \ell}$
		\FOR{$i$ from 1 to $n$}
		\STATE Compute all $z^{(\ell)}$ and $a^{(\ell)}$.
		\STATE $\delta^{(n_\ell)} = -(y-a^{n_{\ell}})\odot \sigma(z^{(n_\ell)})$
		\FOR{$\ell$ from $n_\ell-1$ to 2}
		\STATE $\delta^{(\ell)}=W^{\ell} \delta^{(\ell+1)}\odot \sigma'(z^{(\ell)})$
		\ENDFOR
		\STATE $\nabla_{W^{(\ell)}}C = \delta^{(\ell+1)}{a^{(\ell)}}^T$
		\STATE $\nabla_{b^{(\ell)}}C = \delta^{(\ell+1)}$
		\ENDFOR
	\end{algorithmic}
\end{algorithm}

Algorithm~\ref{alg:bp} is most directly applicable to fully-connected neural
networks.  For structured data, however, convolutional neural networks (CNNs)
are more useful.  The fundamental
unit of computation in a CNN are convolutions -- which are stored as arrays of
some dimension -- unlike vectors in a fully-connected neural network as
described above. Each neuron in a convolution layer considers input from a
small window (such as \texttt{3x3,5x5}) in an image, applies a convolution and
computes a value.  The computation can be reduced to a matrix-vector
multiplication with redundant weights, allowing the above algorithm to be
applied.

%

\section{Elements of Deep Learning Algorithms for Fault Tolerance Consideration}
\label{sec:motivation}

In this section, we present the motivation of our work. Specifically, we
consider the properties of DL algorithms, distinguishing between MLPs, CNNs and
RNNs in terms of their expected execution on large scale systems. This
distinction provides the necessary guidelines for requirements from MPI in
terms of fault tolerance. As pointed out by Gropp and
Lusk~\cite{gropp:ijhpca02}, "fault tolerance is a property of MPI programs and
specification". Hence, it is critical to consider these in conjunction.  Our
first element of discussion is the expected type of parallelism for scaling out
DL algorithms.

\subsection{Master-Slave Paradigm}
Over the last few years, several researchers have considered the possibility of
scaling out DL algorithms~\cite{NIPS2012_0598,geeps,projectadam,mxnet,caffeonspark}. The classical work in scaling out DL algorithms
considered a {\em master-slave} paradigm, which was proposed under the
DistBelief framework~\cite{NIPS2012_0598}. It considered a hierarchical organization of
{\em parameter servers} which would hold the latest copy of the model. The
{\em workers} would periodically update the master with their updates and
request the latest copy of the model. Several extensions to this fundamental
paradigm have been proposed in the literature~\cite{geeps,mxnet,firecaffe}.
The limitations of the master-slave model have been well-studied in the distributed
systems research~\cite{das:arxiv16,lian:nip15}. In addition to being a single point of failure, and a
communication bottleneck, the limitation of this approach is that the
convergence of master-slave paradigm worsens at scale-out. For extreme scale
systems, this approach is infeasible. Hence, we disregard this approach which
would be leveraged in practical deployments especially of HPC systems such as
Leadership Class Facilities (LCFs).

It is also worthwhile noting that this approach is amenable to fault tolerance,
especially if the reliability of the parameter server is higher than workers. A
possible implementation in the master-slave paradigm is either re-spawning of new workers and splitting the original training set among these new workers (by
using \texttt{MPI\_Comm\_spawn}) or continue executing in the presence of faults
using the remaining set of compute nodes. Other researchers have made similar
observations in the context of generic master-slave
applications~\cite{gropp:ijhpca02} and they are readily applicable to DL
algorithms. However, due to the fundamental scaling issues of the master-slave
paradigm for DL implementations, we disregard this approach from
implementation.

\subsection{Model Parallelism}

\begin{figure}[hptb]
\centering
\includegraphics[width=0.7\columnwidth]{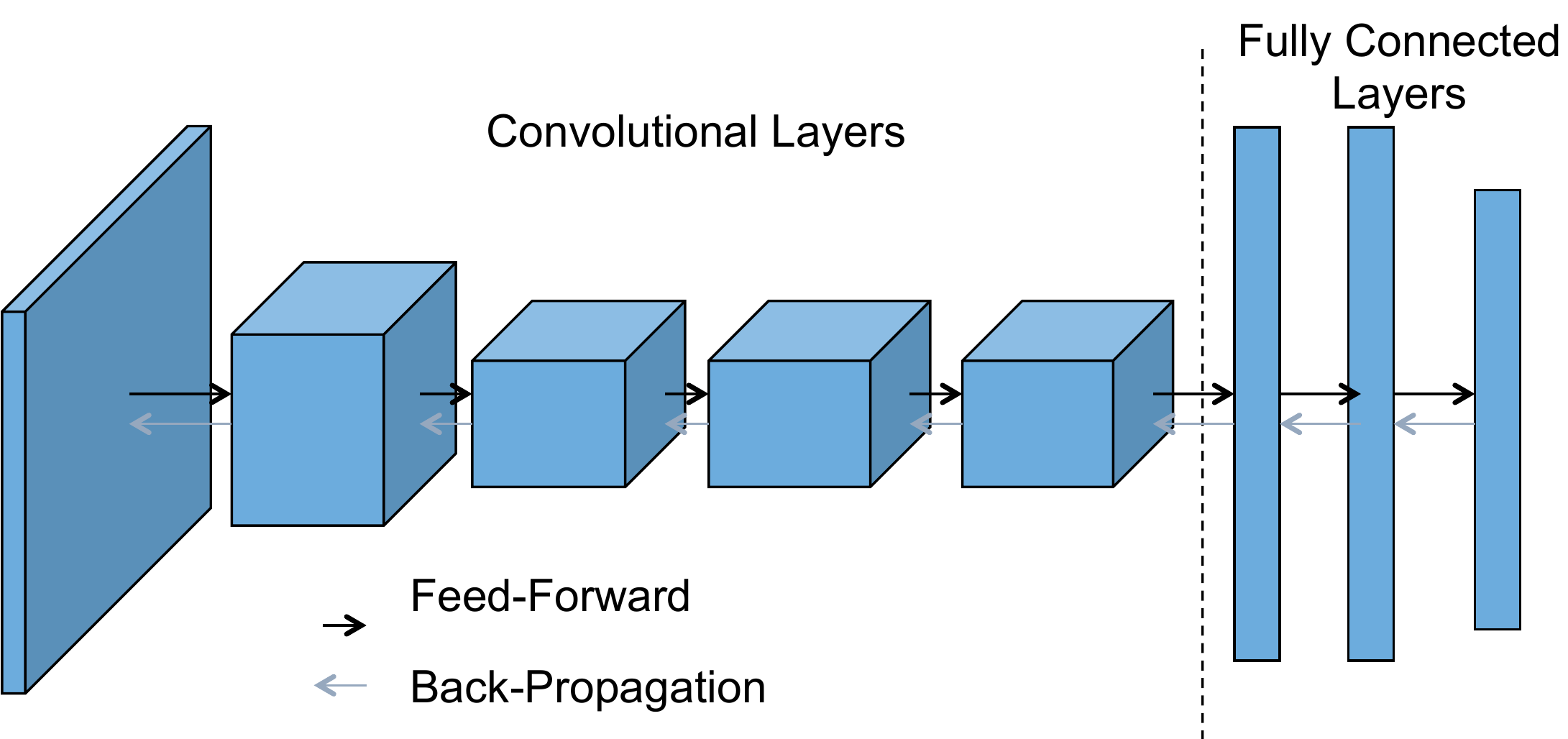}
\caption{A pictorial representation of AlexNet~\cite{imagenet} neural network topology}
\label{fig:alexnet_topology}
\end{figure}

Another possibility which has been presented in literature for scaling out DL
implementations is {\em model parallelism}. In this specific type, individual
layers of the overall DNN model are split among different compute nodes. The
training set itself is split among the compute nodes as well. Let us consider
the example of the AlexNet neural network topology as shown in
Figure~\ref{fig:alexnet_topology}. In a sample execution of model parallelism,
each of the hidden layers is resident on a single compute node.

During the feedforward step, a batch of samples is executed on the first
hidden layer. The output of the first hidden layer -- which is typically
referred to as activations -- is forwarded to the next hidden layer,
resulting in point-to-point communication between two compute nodes. This
procedure is repeated untill the last layer of the DNN is reached, at which point
the {\em error} is calculated.
During the back-propagation step, the error is used to calculate the updates to the
weights ({\em gradients}) which are communicated between compute nodes in the
reverse order to the feedforward step.

\subsection{Data Parallelism}
\begin{figure}[hptb]
\centering
\includegraphics[width=0.7\columnwidth]{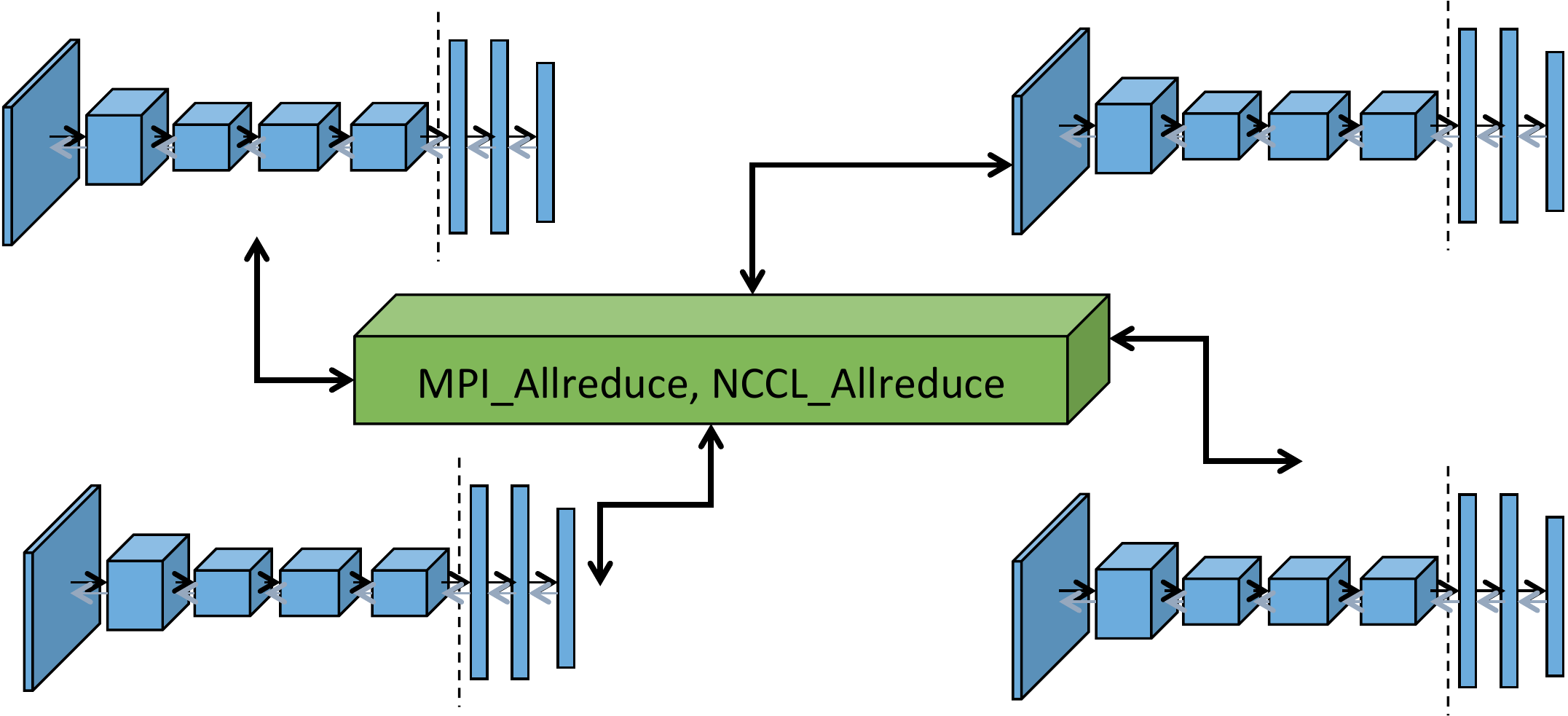}
\caption{A pictorial representation of data parallelism in DL algorithms using
AlexNet neural network topology and four compute nodes. The model is
synchronized at the end of each batch using
\texttt{MPI\_Allreduce} and other primitives such as NVIDIA Collective Communication Library (NCCL)}
\label{fig:data_parallelism}
\end{figure}

A widely used option in scaling out DL implementations is {\em data
parallelism}~\cite{zheng:icpads16,siegel:bigdata16,siegel2016adaptive,vishnu:cluster15-a,vishnu:cluster15-b,vishnu:ipdps16,shohdy:icpp16,shohdy:hipc16,zheng:icpads16}. Under this type of parallelism, the model is replicated and the
data is split among multiple compute nodes. A pictorial representation of the
data parallelism is shown in the Figure~\ref{fig:data_parallelism}.

As shown in the figure, at the end of each batch each compute node (assuming
that the implementation uses shared address space programming model such as
OpenMP/pthread on a node) executes an \texttt{MPI\_Allreduce}. By executing the
all-to-all reduction primitive, the algorithm ensures that it is equivalent to
the default sequential DL algorithm~\cite{Keuper:MHLPC2015}.  An important
consideration for data parallelism is that strong scaling of work is essential
to ensure the equivalence of the implementation to the sequential algorithm.
Specifically, let us consider a batch size $b$, and let $n$ be the number of
compute nodes. The overall expected complexity of the data parallelism based
implementation is $\Theta\left(\frac{b}{n} + \log(n)\right)$. Naturally, the
ratio of communication to computation increases with strong scaling -- which is
a potential downside to data parallelism. Several solutions have been proposed
to handle this situation~\cite{Krizhevsky14oneweird}. One possibility is to
consider increasing the batch size and increasing the values of other
hyper-parameters (such as learning rate) by
Krizhevsky~\cite{Krizhevsky14oneweird}. Recently proposed solutions such as
S-Caffe~\cite{scaffe} improve the scalability of data parallelism by leveraging
the overlap of communication with computation. While a few of these
approaches provide strict
equivalence to the sequential algorithm, other approaches such as
asynchronous variants (also referred to as asynchronous gradient descent (AGD))
are still useful, but do not provide strict equivalence to the default stochastic gradient descent (SGD)
algorithm.

\subsection{Scalability Analysis}
An advantage of model parallelism is its potential to scale-out the DL
algorithms very well. For example, if there are 1,000 layers, then in an ideal
situation each compute node may have one layer, resulting in scale-out.
However, there are several reasons to not consider pure model parallelism based
techniques for scaling out the DL implementations.

\begin{table}[!t]
\centering
\begin{tabular}{|c|c|c|c|c|}
\hline
& Method & Symbol\\
\hline
1 & Features in previous layer          & $n_1$\\
2 & Features in current layer           & $n_2$\\
3 & Activation shape in previous layer  & $x_1\times x_2$ \\
4 & Window size in current layer        & $w_1\times w_2$    \\
5 & Strides for current layer           & $s_1\times s_2$    \\
\hline
\end{tabular}\\
\caption{Symbols Used For Computing Activations and Parameters}
\label{table:activations}
\end{table}

For any DNN with several hidden layers, let us consider two consecutive layers:
$L_1$ and $L_2$.  We compute the number of parameters and activations for $L_2$
as follows:
If $L_2$ is a convolution layer, then the array containing the parameters is
$w_1\times w_2\times n_1\times n_2$.  However, the activations are an array of
size $\frac{x_1}{s_1}\times \frac{x_2}{s_2}\times n_2$.  As
$w_i<\frac{x_i}{s_i}$, the number of parameters is less than the number of
activations.  Conversely, if $L_2$ is a fully-connected layer, then there are
$n_1\times n_2$ parameters and only $n_2$ activations.
We note that if $L_1$ is a convolution layer, then $n_1$ must be replaced by
the total number of activations of that layer, namely $\frac{x_1}{s_1}\times \frac{x_2}{s_2}\times n_1$.
This implies, in
general, convolutional networks have lesser communication volume of weights
than activations. However, for fully-connected networks the trend is reversed.
In the case of AlexNet (shown in Figure~\ref{fig:alexnet_topology}), a
well-established DNN architecture, this can be seen directly.  The first
convolutional layer contains 34,848 parameters, but has 301,056 activations, a
difference of an order of magnitude.  The final fully-connected layer, however,
has 4,096,000 parameters and 1,000 activations.

\subsection{Lessons Learned}
In the previous section, we provided an in-depth discussion on the
possibilities of scaling out DL implementations.  While model parallelism looks
attractive, in practice the ratio of communication of activations to model
parameters prohibits effective scaling.  This is because CNNs are increasingly
becoming commonplace -- including the winners of last 5 years of ImageNet
classification challenge~\cite{imagenet}.  At the same time, data parallelism
provides scaling out possibilities, but has limitations regarding the growth of
batch size~\cite{Keuper:MLHPC2016}. However, solutions proposed by other
researchers~\cite{Krizhevsky14oneweird,scaffe} still make data
parallelism an amenable choice for scaling out DL implementations.  It is worth
noting that a possibility which combines DL model and data parallelism --
hybrid parallelism -- has been proposed in literature as
well~\cite{das:arxiv16}.  However, usually model parallelism is applied on
multiple GPUs/multiple sockets on each compute node and data parallelism is
applied for multiple compute nodes. In essence, we already consider hybrid
parallelism, while implicitly leveraging model parallelism within a node. {\bf
Hence, we consider data parallelism for fault tolerance considerations in rest
of the paper.}

\section{Solution Space}
\label{sec:design}
In this section, we present a solution space for designing fault tolerant DL
implementations using data parallelism, as discussed in the previous section.
An important consideration is exploring the suitability of existing features in
MPI for this purpose, with detailed considerations for the primary proposals.

\subsection{Critical Data Structures in DL Algorithms}
The first step is the identification of critical data structures in DL
algorithms. Specifically, there are several data structures which are used
during the feed-forward and the back-propagation phase of the DL
implementations. During the feed-forward step, the input dataset and the model
are used -- both of which are read-only during the step. However, at the
back-propagation step, the model weights are updated while the input dataset
remains untouched. Hence, the critical data structure for DL implementations is
the model weights which are updated iteratively till convergence. It is worth
noting that there are auxiliary data structures which are updated as well. As
an example for DL implementations with {\em momentum}, data structures such as
{\em history} are updated, which can be recalculated from the model weights. The
gradients -- iterative updates to the model weights -- are calculated during
the back-propagation step. However, they are accumulated iteratively in the
model weights, and hence they are not critical.

\subsection{Process Recovery Model}
An important design point is the process recovery model. Several scientific
applications such as LULESH~\cite{IPDPS13:LULESH} and NAS Parallel
Benchmarks~\cite{Bailey:SC91} typically require a fixed
topology (such as a square/quadratic) in terms of number of MPI processes.
However, there are no such requirements for DL implementations. Hence, it is
possible to continue execution with the remaining set of compute nodes, without
affecting the correctness of the DL implementation. A natural advantage of this
approach is that it requires little support from process managers for practical
deployments. Hence, we use this approach for designing fault tolerant DL
implementation.


\section{Suitability of Existing MPI Features and Proposals}
\label{sec:existing_proposals}
\subsection{Suitability of Existing Error Handling in MPI}
An important design consideration is the suitability of existing error handling
in MPI for designing fault tolerant DL implementations. Specifically, by
initializing with \texttt{MPI\_ERRORS\_RETURN}, it is
possible for a DL implementation to prevent aborting when a fault is detected
either during MPI communication or an external system software component (such
as SLURM).

By setting an explicit error handler, it is possible for the DL implementation
to checkpoint their critical data structures, exit the application and re-start
the application from the recent saved checkpoint. In this specific case, the critical
data structure is the model parameters of the DNN, since the dataset is
read-only, and it can be readily recovered from the disk/file-system. The DL implementation
may be re-started using $n$ compute nodes (if spare compute nodes are
available) or $n - 1$ compute nodes, since DL implementations do not have
specific requirements of a topology.

This approach is definitely a suitable possibility. However, it may not be necessary,
since this will result in a recovery complexity of $O(n)$, a
function of number of compute nodes, instead of the degree of failure. The
reasoning behind this time complexity is due to the fact that the $n - 1$
compute nodes would need to read the entire dataset back from the disk
(prohibitive data movement), in addition to the checkpointed model files.
Hence, this approach may be considered as the baseline approach, but not
necessarily as the optimal approach for handling permanent node faults in DL
implementations.

\subsection{Suitability of User-Level Fault Mitigation (ULFM) Proposal}
One of the fault tolerance proposal which has been
considered for inclusion in the MPI specification for the last few years is
ULFM. The salient
features of ULFM are: 1) ability to provide non-collective global fault
notification, 2) ability to recover from faults by fixing the {\em broken}
communicator on the fly and 3) support for fixed/shrinking process set.

ULFM is particularly suited for applications which have small process-specific
state information. Usually, resetting the global state information is
non-trivial and requires writing a complex error handler. Naturally, for
large-scale applications -- which have been developed over decades -- writing a
correct error handler even for a subset of fault cases is non-trivial.

However, with data parallelism the overall state information that is required
for DL implementations is minimal. Since the model is replicated across the compute
nodes, the DL implementation requires no checkpointing. Hence, ULFM is
potentially the right fit for implementing fault tolerant DL algorithms.

\subsection{Suitability of Reinit Fault Tolerance Proposal}

Recently, Laguna {\em et al.} have proposed {\em Reinit} proposal for handling
faults in MPI.  The objective of Reinit is  to address the limitations of ULFM,
and is particularly suitable to applications where the code complexity of
the recovery module is high. The salient features of the Reinit proposal are: 1)
automatic re-initialization of MPI after a fault is detected, 2)
semi-automatic recovery from the intermediate checkpoints, and  3) ability to handle
shrinking/fixed process set.

We consider the suitability of the Reinit proposal to data parallelism based DL
implementations.  We observe that DL implementations would be required to
check-point the model weights periodically, which would be used by the Reinit
implementation during recovery.  We also observe that the application would be
re-started requiring the entire dataset to be read from the disk.

\subsection{Lessons Learned}
We observe that existing local fault notification in the MPI specification and
implementations may be used for developing fault tolerant DL implementations.
However, there is a significant amount of work needed within MPI and at the
application level (such as intermediate checkpointing) to leverage the existing
functionality. The Reinit proposal is suitable as well. However, there are two
potential downsides that are readily observed: 1) Reinit would require DL
implementations to consider intermediate check-pointing, when the DL algorithm
does not mandate it, and  2) Reinit would require application to read the entire
dataset from the disk, when reading the data could be fairly localized to the
degree of failure.

ULFM has positive attributes which are definitely suitable for designing fault
tolerant DL implementations. The primary functionality that is required is
automatic fixing of the communicator, and reading the partial dataset from the
disk, while continuing to execute with the existing set of compute nodes. The
ULFM specification itself has a few implementation caveats, including the cost
of fault detection (which is relatively lesser for Reinit), cost of global
notification and cost of recovering the communicator. Yet, the overall cost of
computation recovery is at most one batch -- while in the case of Reinit, it is
expected to be much higher depending upon the degree of checkpointing. Hence,
we consider ULFM for implementing fault tolerant DL algorithms. In the next
section, we present the implementation details for ULFM based DL algorithm.

\section{Design and Implementation Details}
\label{sec:implDetail}
\begin{figure}[hptb]
\centering
\includegraphics[width=0.8\columnwidth]{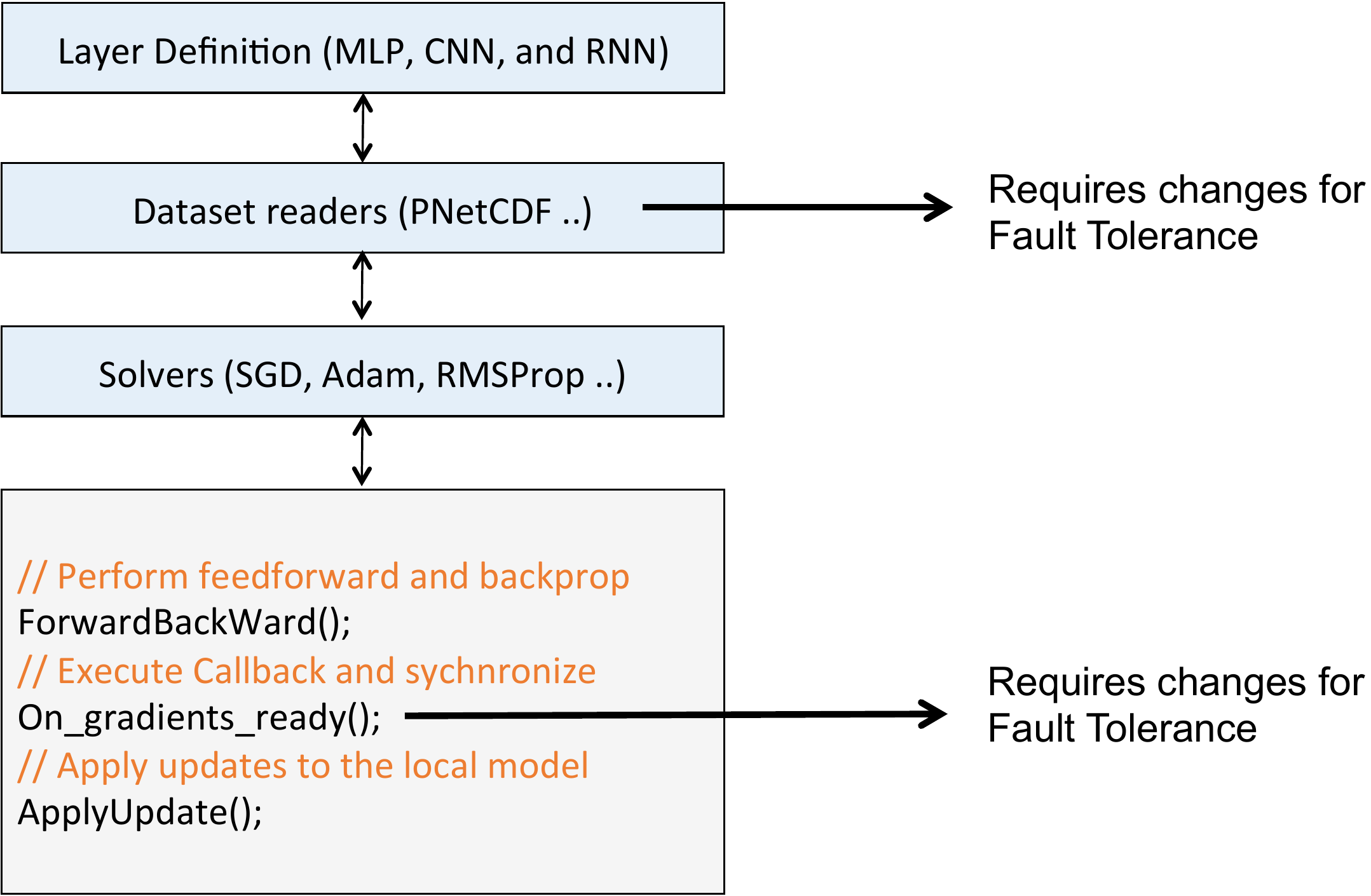}
\caption{Overall Caffe architecture and code flow for the original code. We require changes to dataset reading and the callback for implementing fault tolerance.}
\label{fig:code_snippet}
\end{figure}

In this section, we present design and implementation details for fault
tolerant DL algorithms. We leverage the ULFM implementation provided by OpenMPI
for this purpose and implement our changes in Caffe runtime.
Figure~\ref{fig:code_snippet} shows the overall interaction between different
components.

As shown in figure, the Caffe architecture has layers such as for MLPs, CNNs
and RNNs, which are defined in a prototxt file. Hence, these are already
resident on disk. Our extensions of Caffe -- also referred to as MaTEx-Caffe
for rest of the paper -- support parallel NetCDF format which requires changes
for fault tolerance. Caffe runtime supports several types of solvers such as
SGD, Adam and others. These solvers use a common substrate for data
parallelism, where \texttt{ForwardBackward()} step computes the forward and
back-propagation steps of the overall implementation. The resulting gradients
are then synchronized using a callback, which is extended by MaTEx-Caffe for
using \texttt{MPI\_Allreduce}. The resulting synchronized gradients are then
applied using the \texttt{ApplyGradients} function before the next batch of
samples are ingested.

\subsection{Pseudo-code Walk-through}
\begin{figure}[hptb]
\centering
\includegraphics[width=\columnwidth]{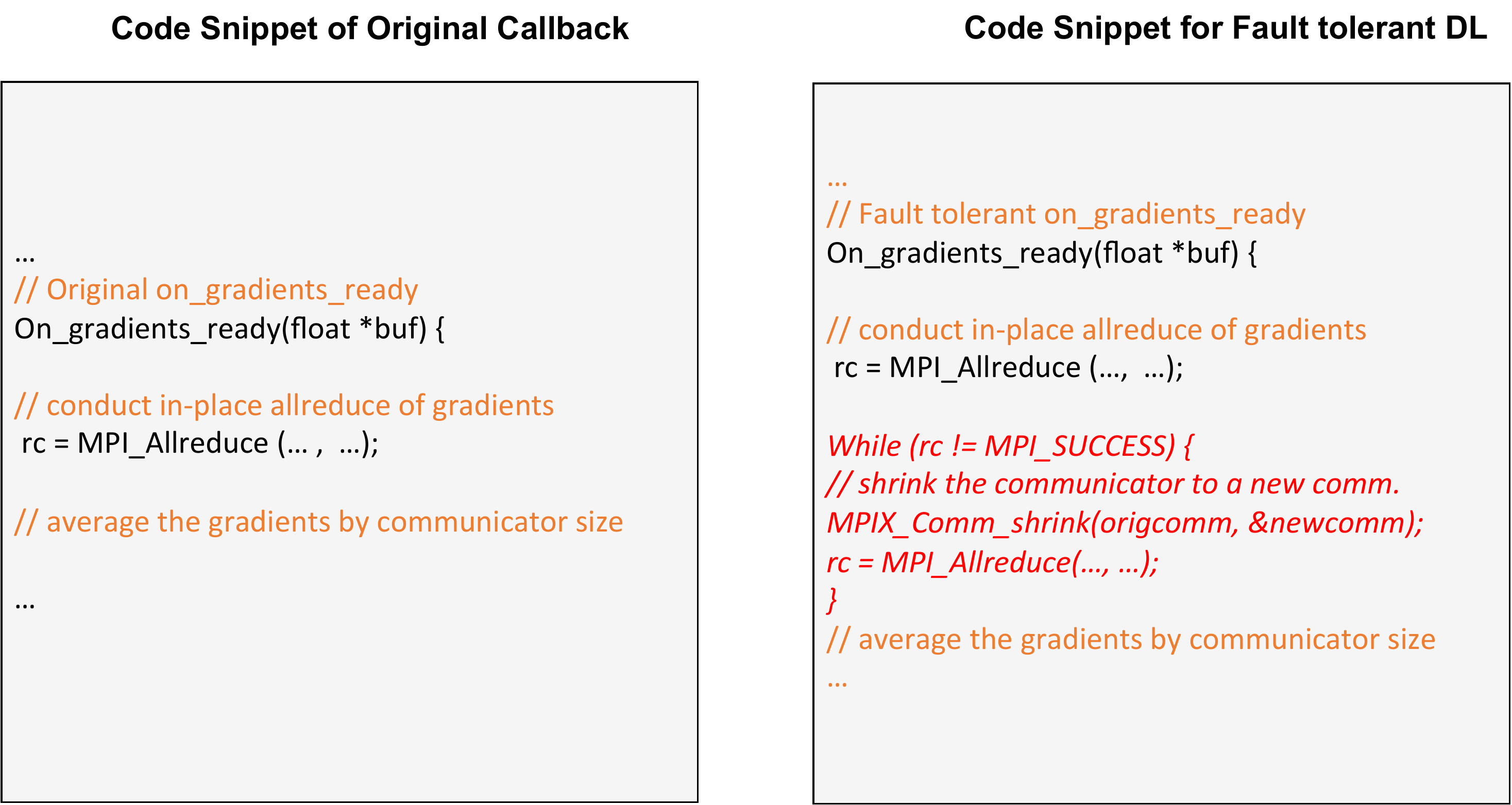}
\caption{Code snippet for ULFM based fault tolerant DL implementation}
\label{fig:ulfm_code}
\end{figure}

Figure~\ref{fig:ulfm_code} shows the difference between the original
non-fault-tolerant implementation of the callback \texttt{on\_gradients\_ready}
and the fault tolerant version on the right. Figure~\ref{fig:code_flow}
presents the code changes for data readback in the fault tolerant version and
compares it to the original code.

\subsubsection{Original Callback}
The original call-back receives the gradients from the \texttt{Forwardbackward}
function and uses an all-to-all reduction to synchronize the gradients
across all compute nodes. The resulting gradient is divided by the number of
compute nodes and applied to the local model using \texttt{ApplyGradients}
function.

\subsubsection{Callback with ULFM Changes}
As shown in Figure~\ref{fig:ulfm_code}, the ULFM changes are handful and
primarily restricted to a single callback. Specifically, when a fault is
detected, we leveraget the \texttt{MPIX\_Comm\_shrink} function is used to shrink the communicator
from original to the new communicator. Once the communicator is reset, then all to all reduction is retried till return code is \texttt{MPI\_SUCCESS}.

\subsubsection{Data Readback for Fault Tolerant Version}
\begin{figure}[hptb]
\centering
\includegraphics[width=\columnwidth]{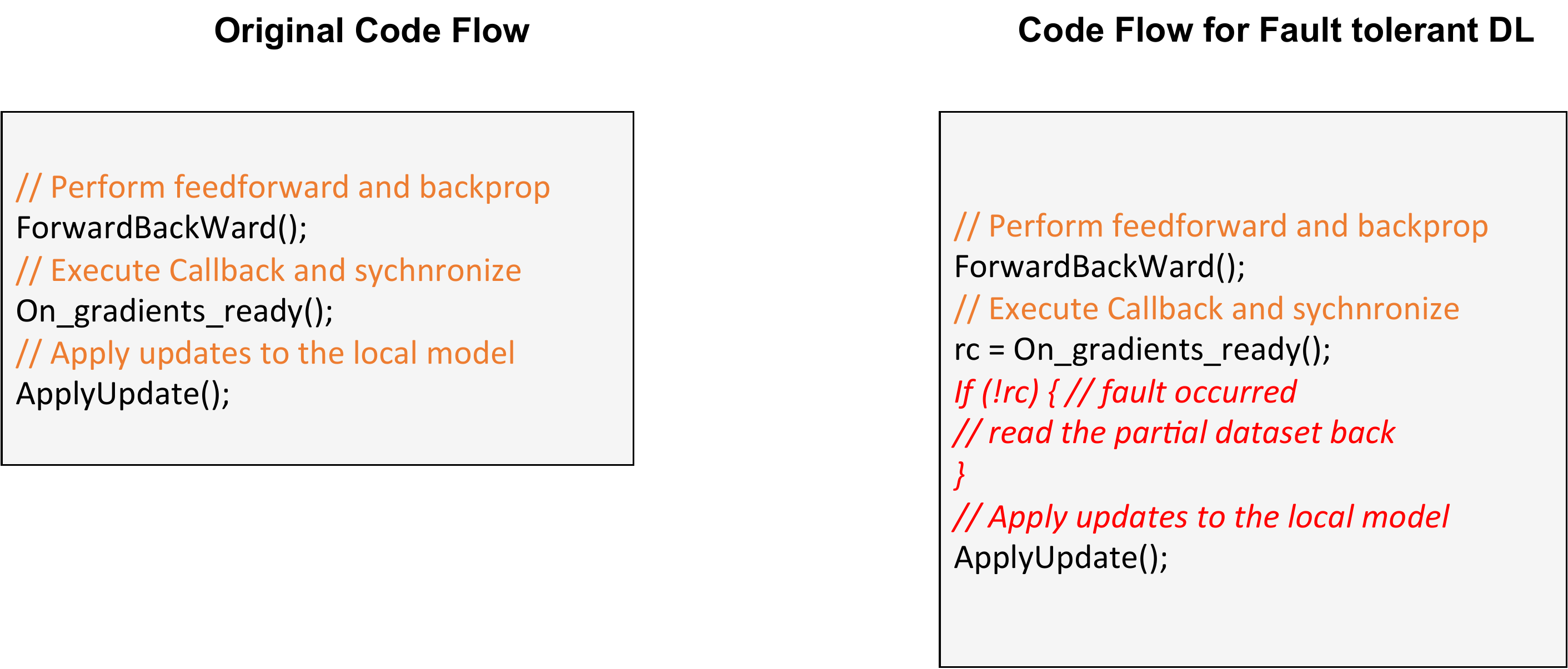}
\caption{Code flow for original and fault tolerant implementations for data
readback}
\label{fig:code_flow}
\end{figure}

Figure~\ref{fig:code_flow} shows the original code flow and the code flow for fault tolerant DL. The data is read only when a fault is detected by the \texttt{on\_gradients\_ready}.


\section{Performance Evaluation}
\label{sec:exp}
In this section, we present a detailed performance evaluation of the proposed
fault tolerant approach using ULFM.  Table~\ref{table:arch} shows the hardware
and software details of our systems.  Table~\ref{table:heur_desc} shows a
description of proposed approaches that we have implemented.
Table~\ref{table:datasets} provides an overview of datasets, and the associated
neural networks.

\begin{table*}[!htbp]
		\centering
		\begin{tabular}{|c|c|c|c|c|c|c|c|c|c|}
				\hline
				Name  & CPU (\#cores) & GPU & Baseline Caffe & Network & MPI & cuDNN & CUDA & Nodes & \#cores\\
				\hline
				\bf{PUMA} & Haswell (20) & N/A & Intel-Caffe~\cite{caffe-intel} & IB &
        ULFM-OpenMPI & N/A &  N/A & 16 & 320\\
				\hline
		\end{tabular}\\
		\caption{Hardware and Software Description. IB (InfiniBand). The
		proposed research extends Baseline-Caffe incorporating
		architecture specific optimizations provided by vendors.}
		\label{table:arch}
\end{table*}
\begin{table*}[!t]
		\centering
        \small
		\begin{tabular}{|c|c|c|l|}
				\hline
				Name  & Type  & Implemented & Description of DL Algorithm and Implementation\\
				\hline
				SGD (Default) & Stochastic Gradient Descent & Yes & Implements strong scaling by dividing batch and all-to-all reduction. \\
				FT-SGD & Fault Tolerant Stochastic Gradient Descent & Yes & Implements strong scaling by dividing batch and all-to-all reduction. \\
				\hline
		\end{tabular}\\
		\caption{Description of Approaches}
		\label{table:heur_desc}
\end{table*}

\begin{table*}[!t]
		\centering
		\begin{adjustbox}{max width=\textwidth}
		\begin{tabular}{|c|c|c|c|c|c|c|c|}
				\hline
				Dataset  & Model     & Description & Training Samples & Validation Samples & Image Size            & Classes \\
				\hline

                               MNIST~\cite{mnistlecun}    & LeNet-3~\cite{lecun1998gradient} & Handwritten Digits & 60000            & 10000          &     $28 \times28$ & 10        \\
                            CIFAR-10~\cite{Krizhevsky09learningmultiple} & Caffe-default  & Small Images & 60000            & 10000              & $32 \times32\times3$  & 10  \\
				ImageNet~\cite{ILSVRC15} & AlexNet~\cite{NIPS2012_4824} & Diverse Images       & 1,281,167          & 50,000              & $256\times256\times3$  & 1,000 \\
				ImageNet~\cite{ILSVRC15} & GoogLeNet & Diverse Images       & 1,281,167          & 50,000              & $256\times256\times3$  & 1,000 \\
				\hline
		\end{tabular}
		\end{adjustbox}
		\caption{Datasets and Models evaluated}
		\label{table:datasets}
\end{table*}


\subsection{Objective}
The objective of our performance evaluation is to understand the performance
overhead of using ULFM based implementation and correctness implications (if
any) of the existing ULFM implementations.

\subsection{Fault Injection Methodology} To emulate the process faults, we
insert a fault in a process by using \texttt{SIGKILL}. For controlled experiment the fault is injected after 300 batches. This
effectively emulates compute node faults, since we use one process on a compute
node and multiple threads for each process.

\subsection{Correctness Analysis}
\begin{figure*}[!htbp]
\centering
\subfloat[AlexNet Loss 4 Nodes]{\includegraphics[width=0.33\textwidth]
          {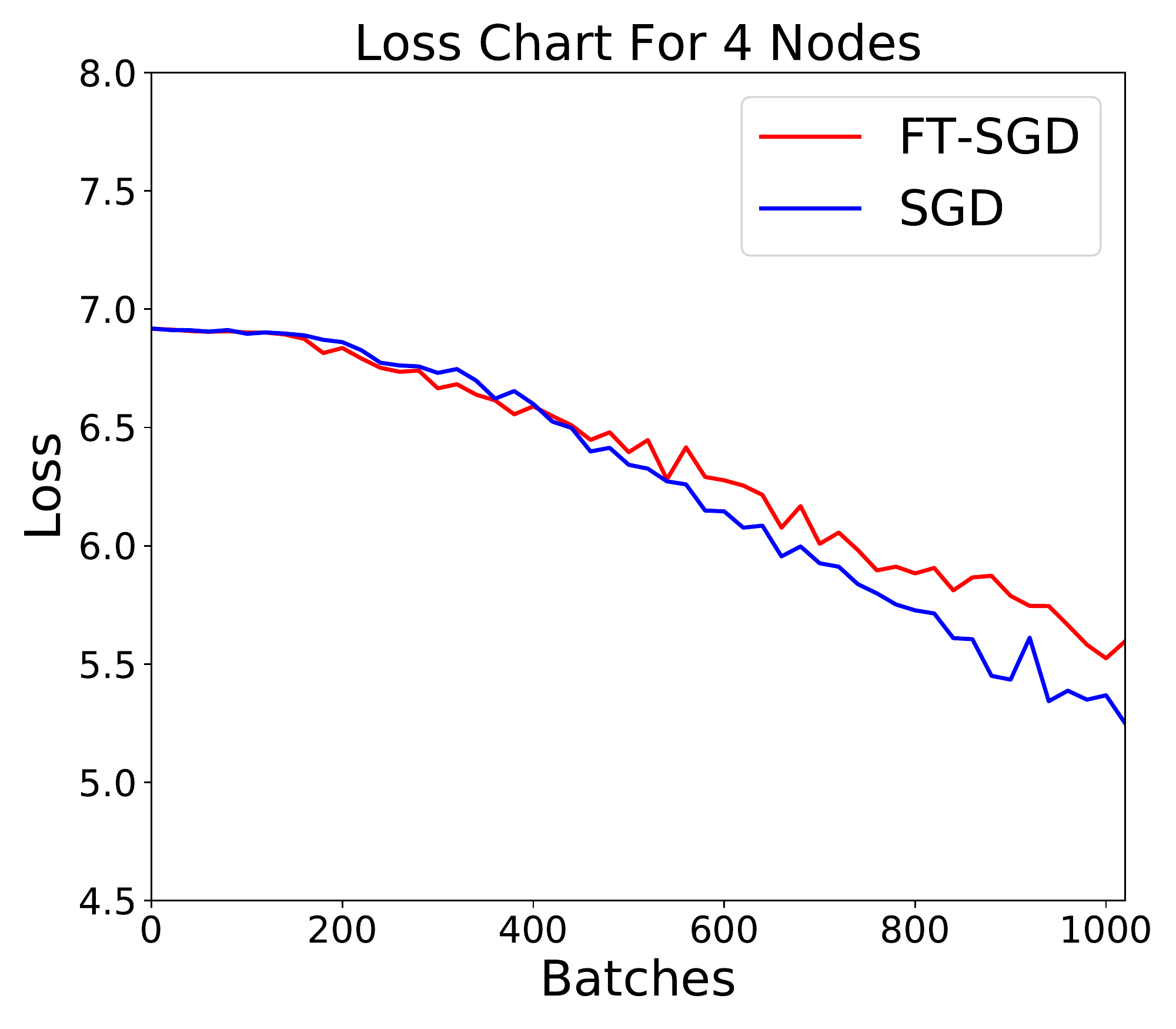}\label{fig:alexnet_loss_n4}}
\subfloat[AlexNet Loss 8 Nodes]{\includegraphics[width=0.33\textwidth]
          {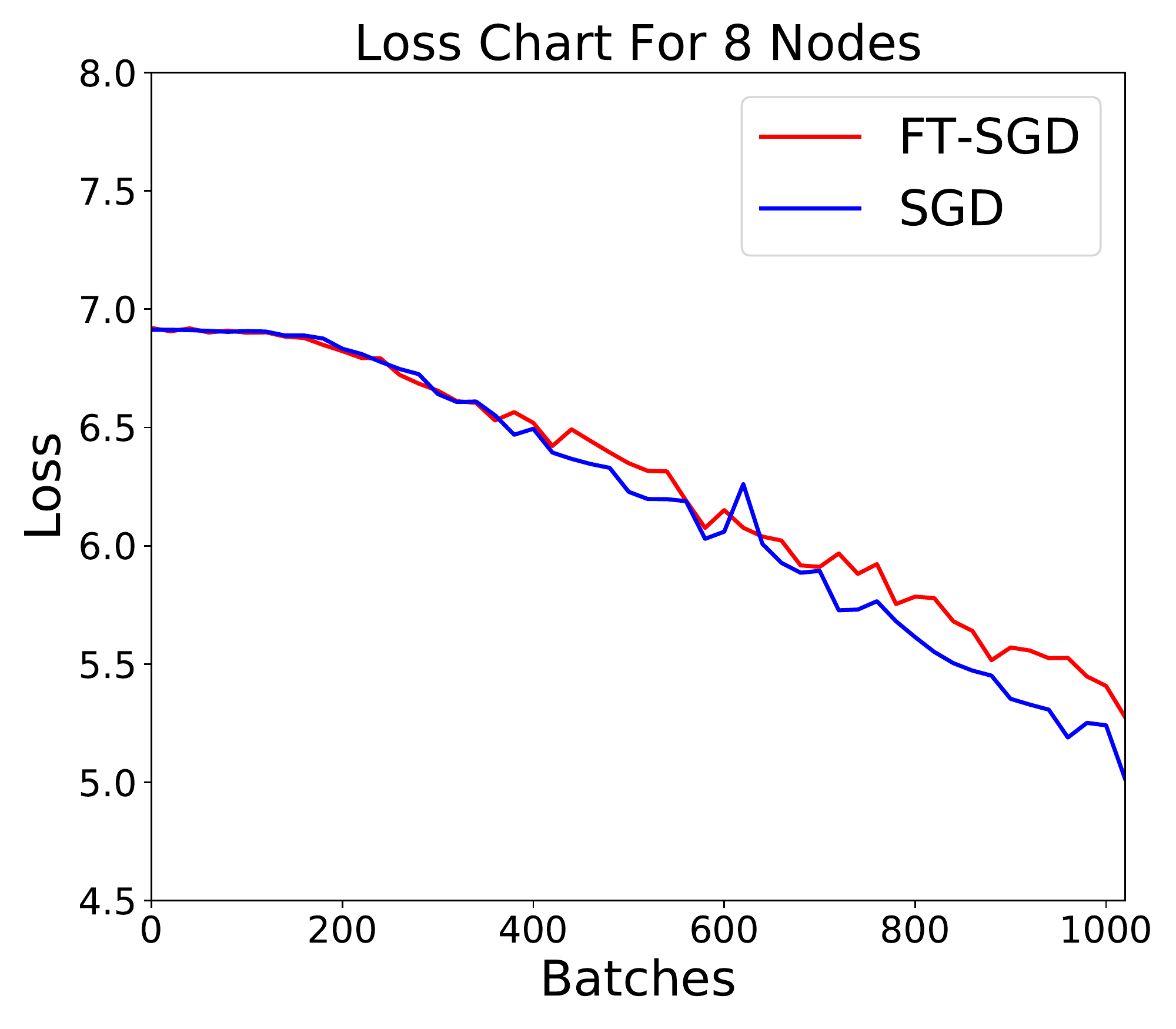}\label{fig:alexnet_loss_n8}}
\subfloat[AlexNet Loss 16 Nodes]{\includegraphics[width=0.33\textwidth]
          {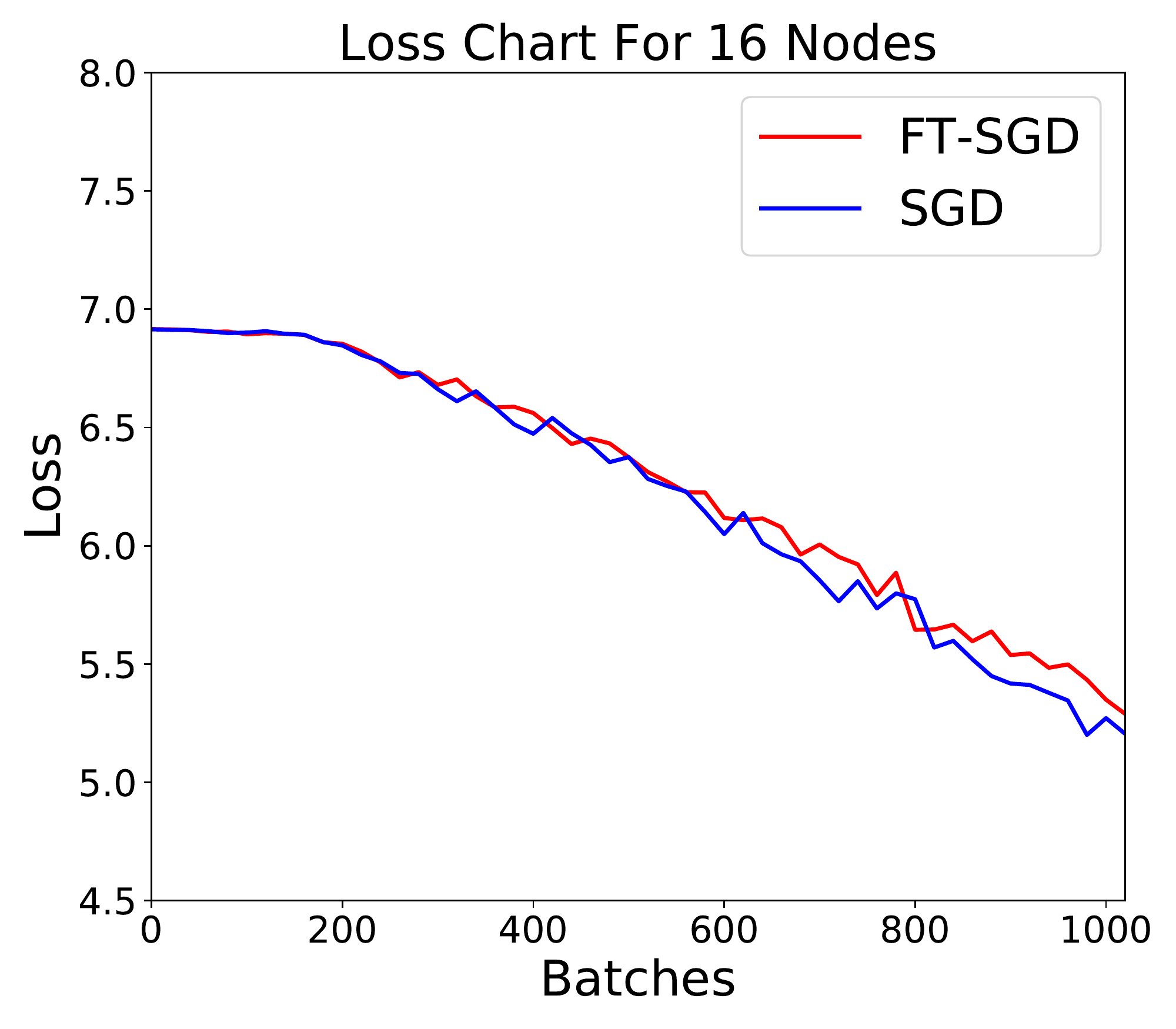}\label{fig:alexnet_loss_n16}}
\\

\subfloat[GoogleNet Loss 16 Nodes]{\includegraphics[width=0.33\textwidth]
          {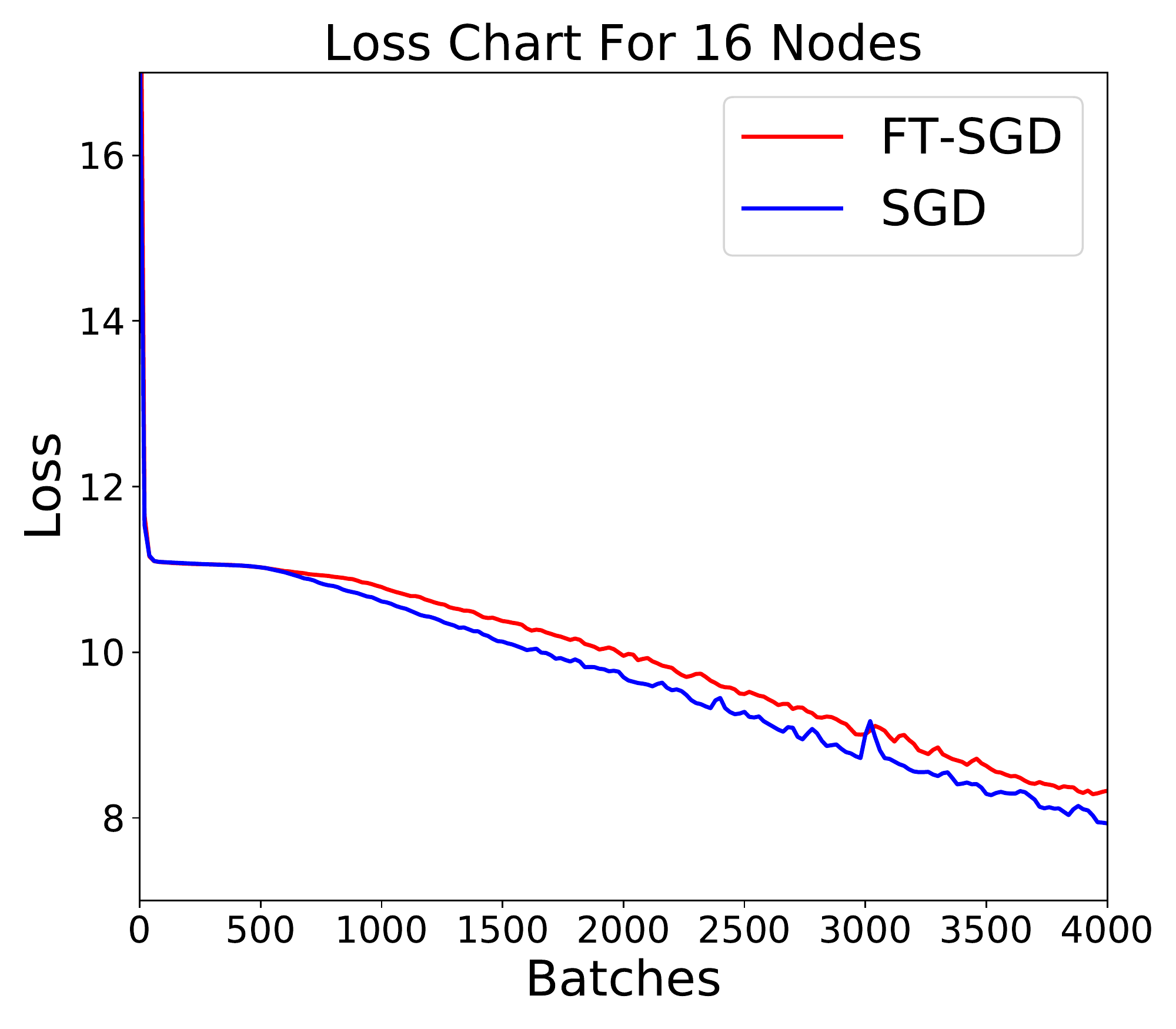}\label{fig:googlenet_loss_n16}}
\subfloat[CIFAR-10 Loss 16 Nodes]{\includegraphics[width=0.33\textwidth]
          {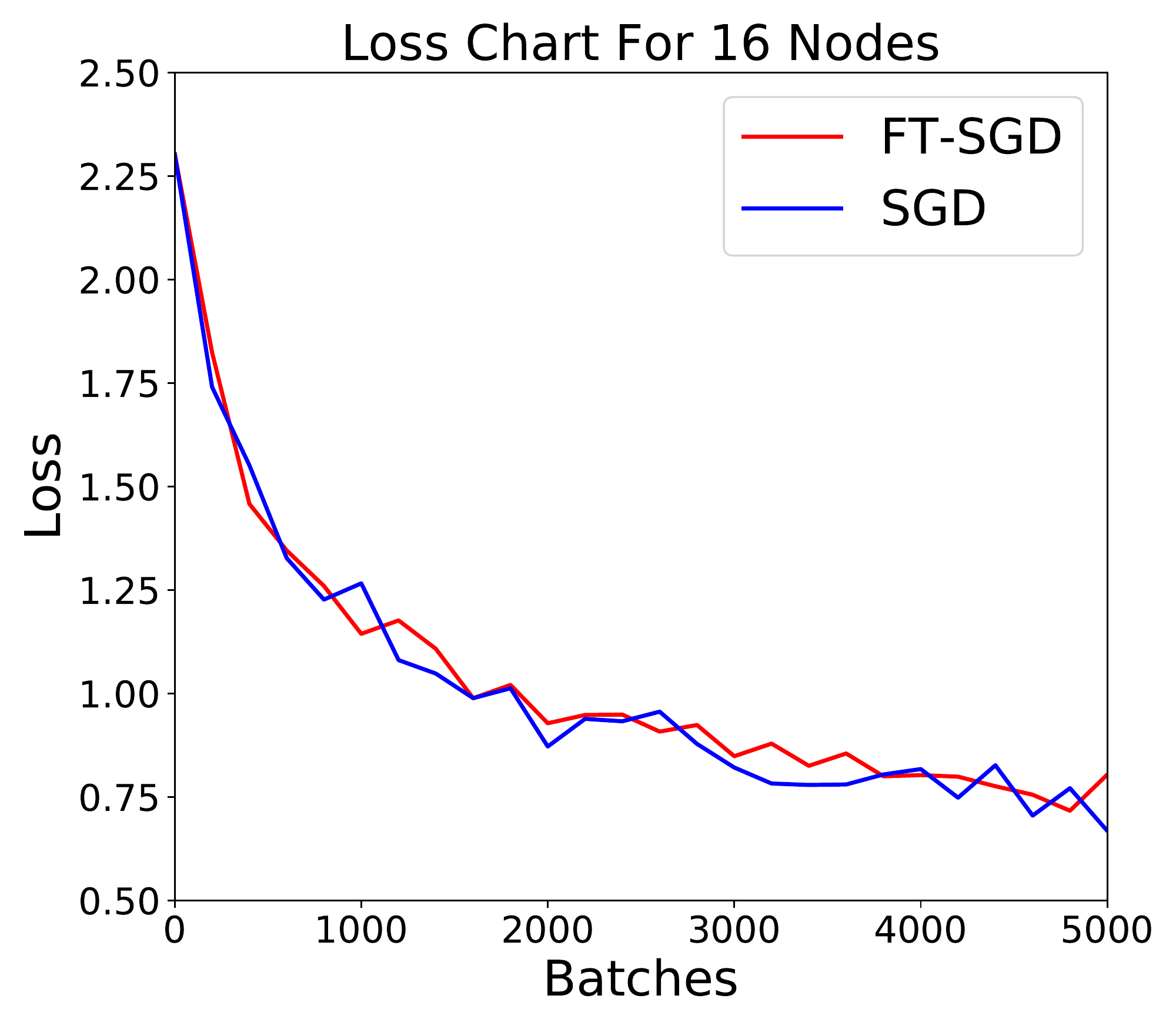}\label{fig:cifar10_loss_n16}}
\subfloat[MNIST Loss 16 Nodes]{\includegraphics[width=0.33\textwidth]
          {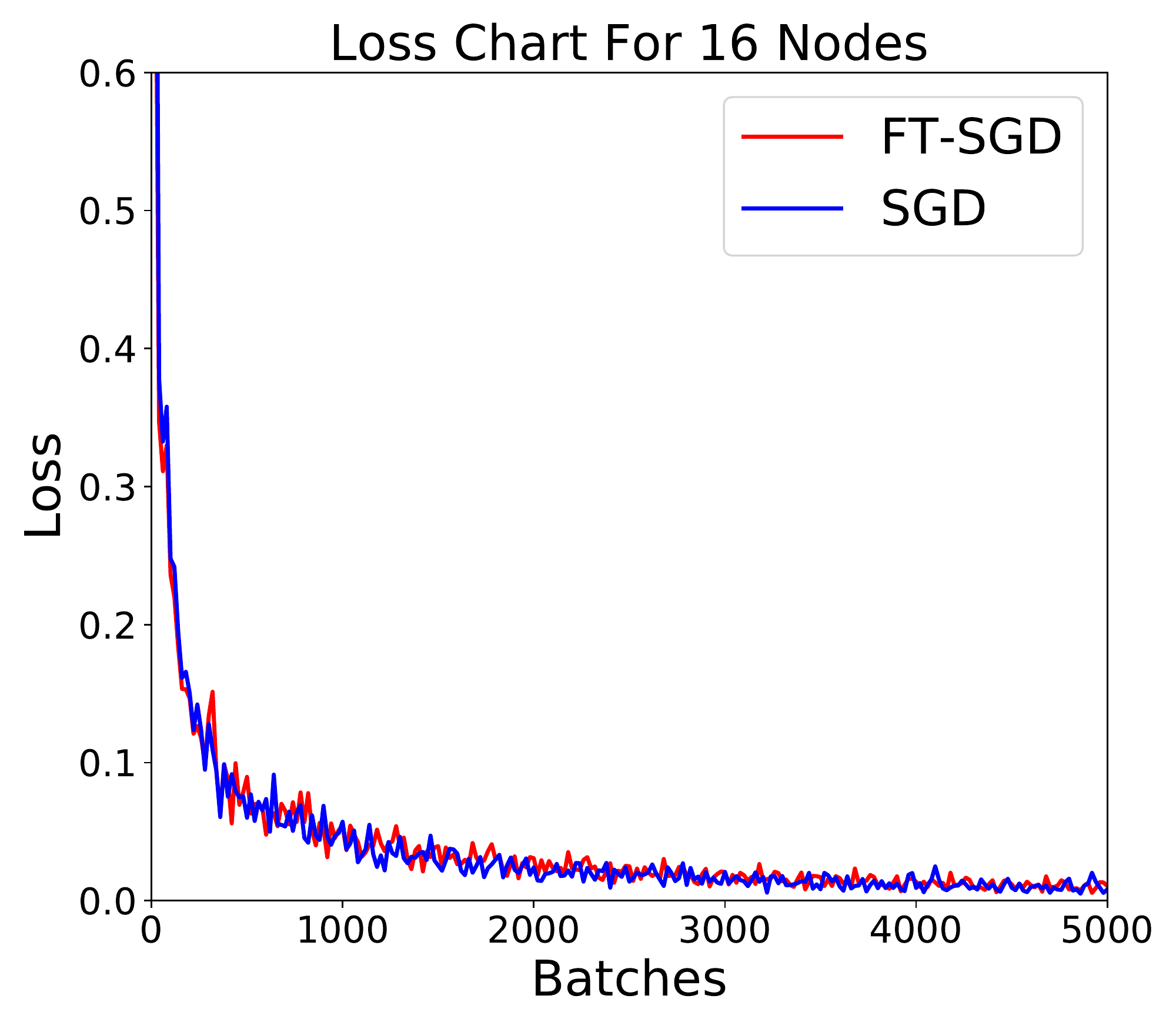}\label{fig:mnist_loss_n16}}
\caption{Loss Charts for Several Datasets comparing SGD and FT-SGD}
\end{figure*}

For understanding the correctness, we compare the {\em loss curves} -- a
measure of the error as observed during the training phase. The curves are
compared for SGD and FT-SGD implementations in
Figures~\ref{fig:alexnet_loss_n4}, ~\ref{fig:alexnet_loss_n8}, and
~\ref{fig:alexnet_loss_n16} using 4, 8, and 16 compute nodes respectively. The
FT-SGD evaluation consists of exactly one process fault -- which is usually the
case in real scenarios. We observe that the loss curves for both SGD and FT-SGD
implementation track each other roughly. However, the FT-SGD is behind since
with reduced number of available compute nodes, the overall batch size is
reduced as well.  The effect is diminished on 16 compute nodes since
the overall effect of losing one compute node is reduced. Similar for other
datasets as shown in Figures~\ref{fig:googlenet_loss_n16},
~\ref{fig:cifar10_loss_n16} and ~\ref{fig:mnist_loss_n16}, the convergence of
FT-SGD and SGD is similar.

\subsection{Performance Analysis} 
\begin{figure*}[!htbp]
\centering
\subfloat[AlexNet Average per Batch Time]{\includegraphics[width=0.49\textwidth]
          {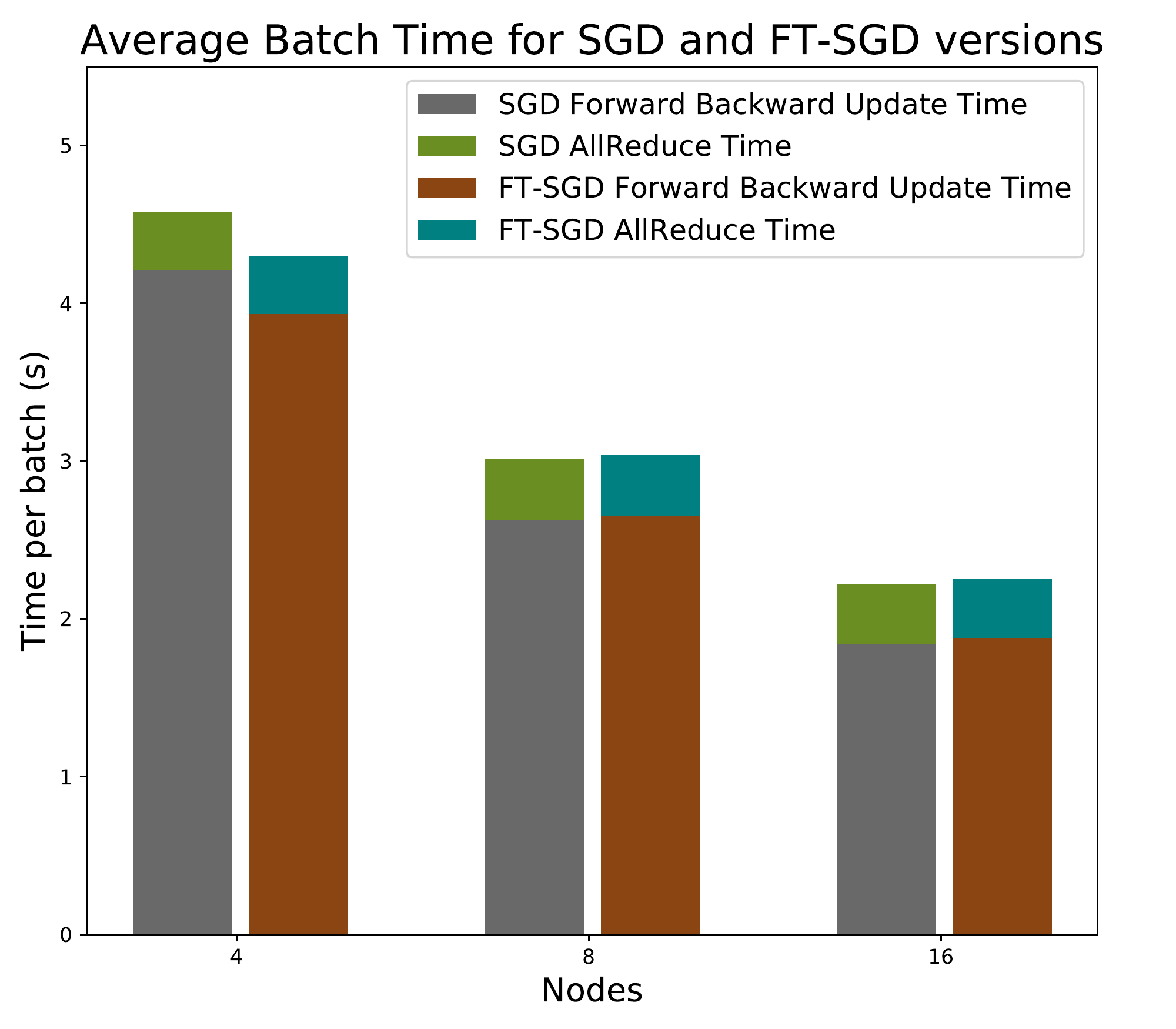}\label{fig:alexnet_run_time}}
\subfloat[GoogleNet Average per Batch Time]{\includegraphics[width=0.49\textwidth]
          {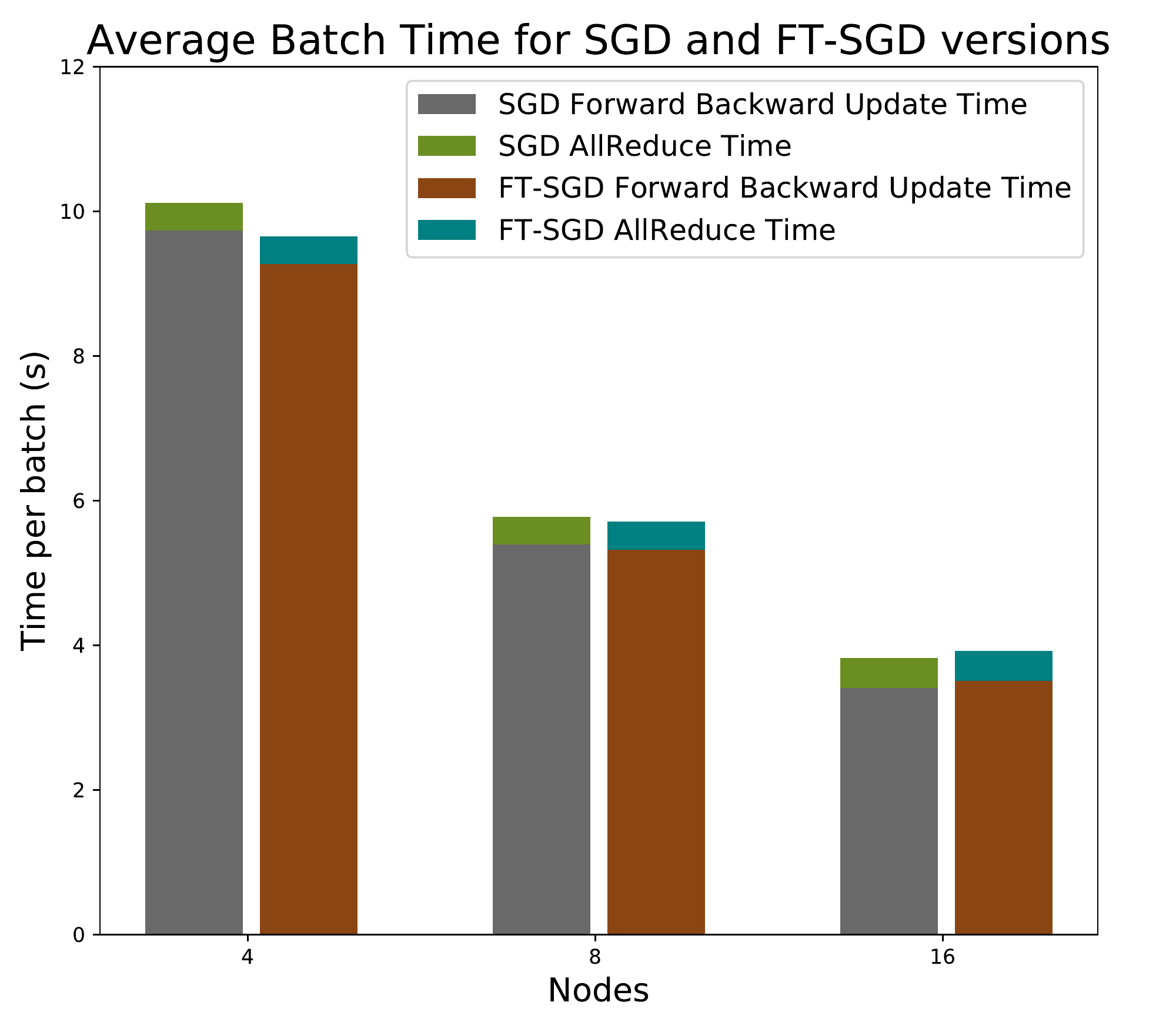}\label{fig:googlenet_run_time}}
\\
\subfloat[CIFAR-10 Average per Batch Time]{\includegraphics[width=0.49\textwidth]
          {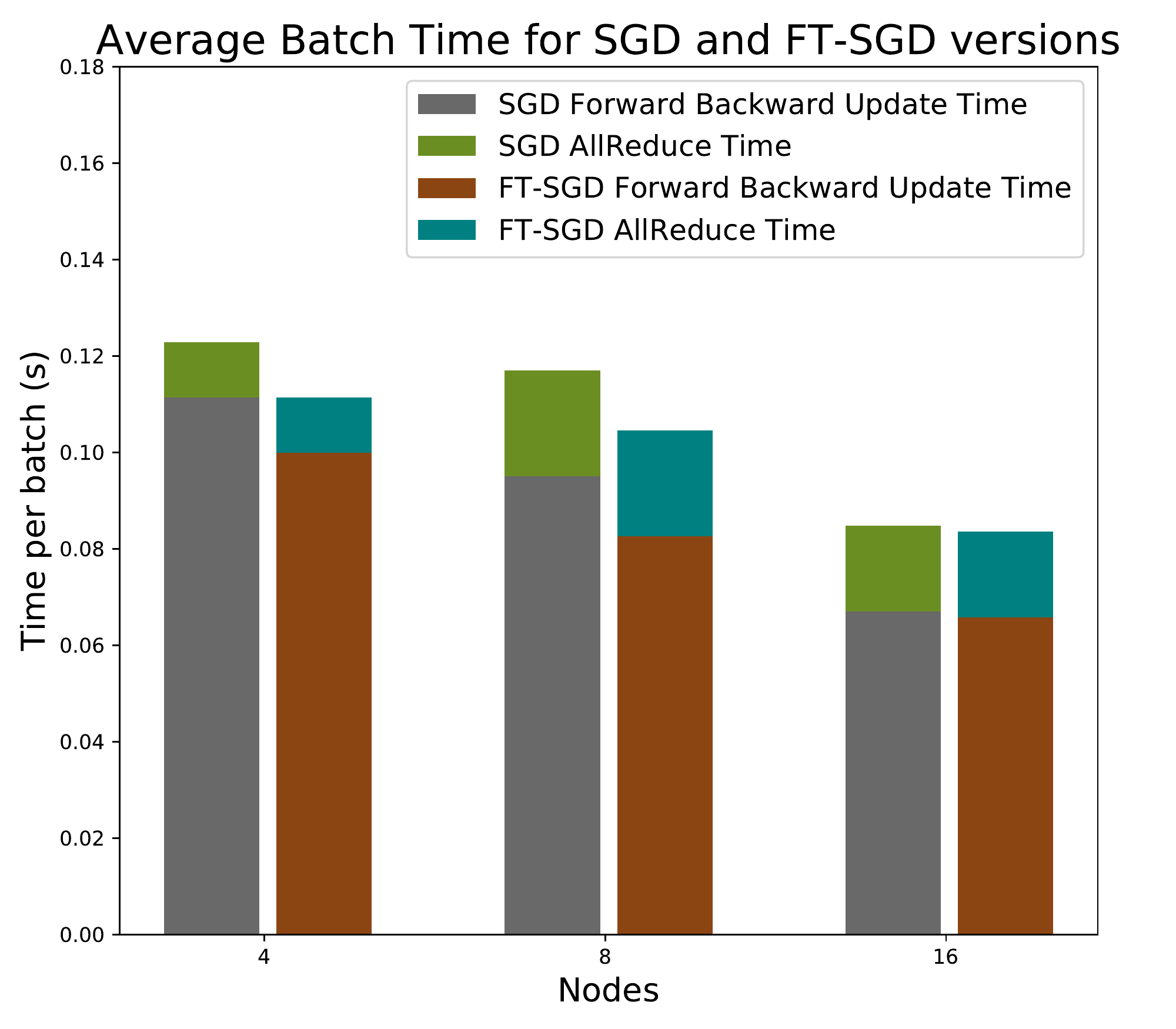}\label{fig:cifar10_run_time}}
\subfloat[MNIST Average per Batch Time]{\includegraphics[width=0.49\textwidth]
          {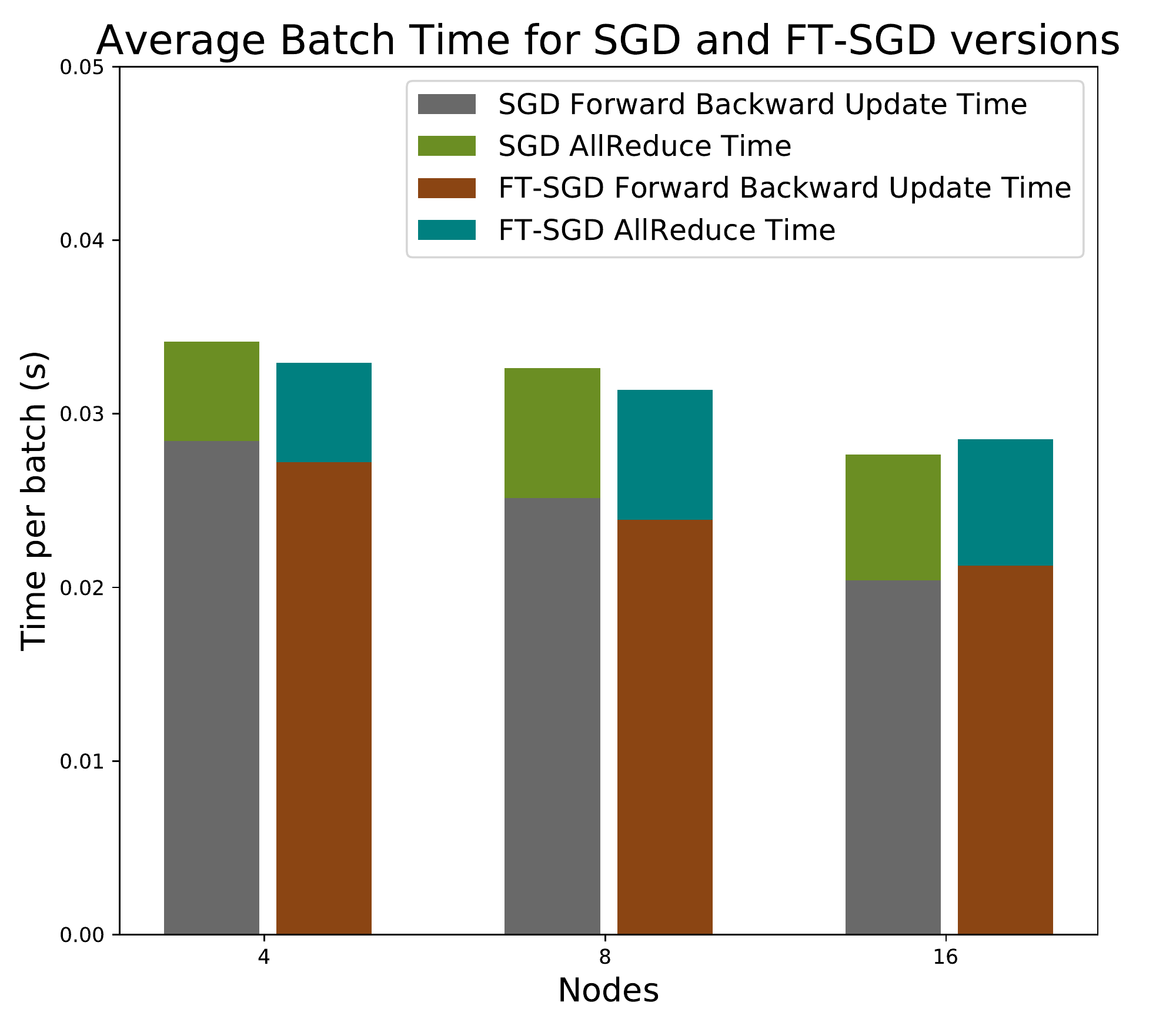}\label{fig:mnist_run_time}}
\caption{Performance Comparisons of SGD and FT-SGD}
\end{figure*}

Figures~\ref{fig:alexnet_run_time},
~\ref{fig:googlenet_run_time}, ~\ref{fig:cifar10_run_time}
and~\ref{fig:mnist_run_time} shows the performance comparisons of SGD and
FT-SGD using AlexNet, GoogLeNet, CifarNet (CIFAR10) and LeNet-3 (MNIST). The
overall evaluation uses 1024 batches, which is a relatively small number of
batches in comparison to the standard number of batches such as 60K for
AlexNet. For the FT-SGD evaluation, exactly one process executes
\texttt{SIGKILL} after 300 batch updates, resulting in $n - 1$ remaining number
of compute nodes. Since the batch on each compute node remains constant, the
overall computation time as observed on each compute node is similar for FT-SGD
and SGD. Also, since we have one compute node failure, the overall difference
in communication time is also negligible.

\begin{figure*}[!htbp]
\centering
\subfloat[AlexNet Data Load Times]{\includegraphics[width=0.25\textwidth]
          {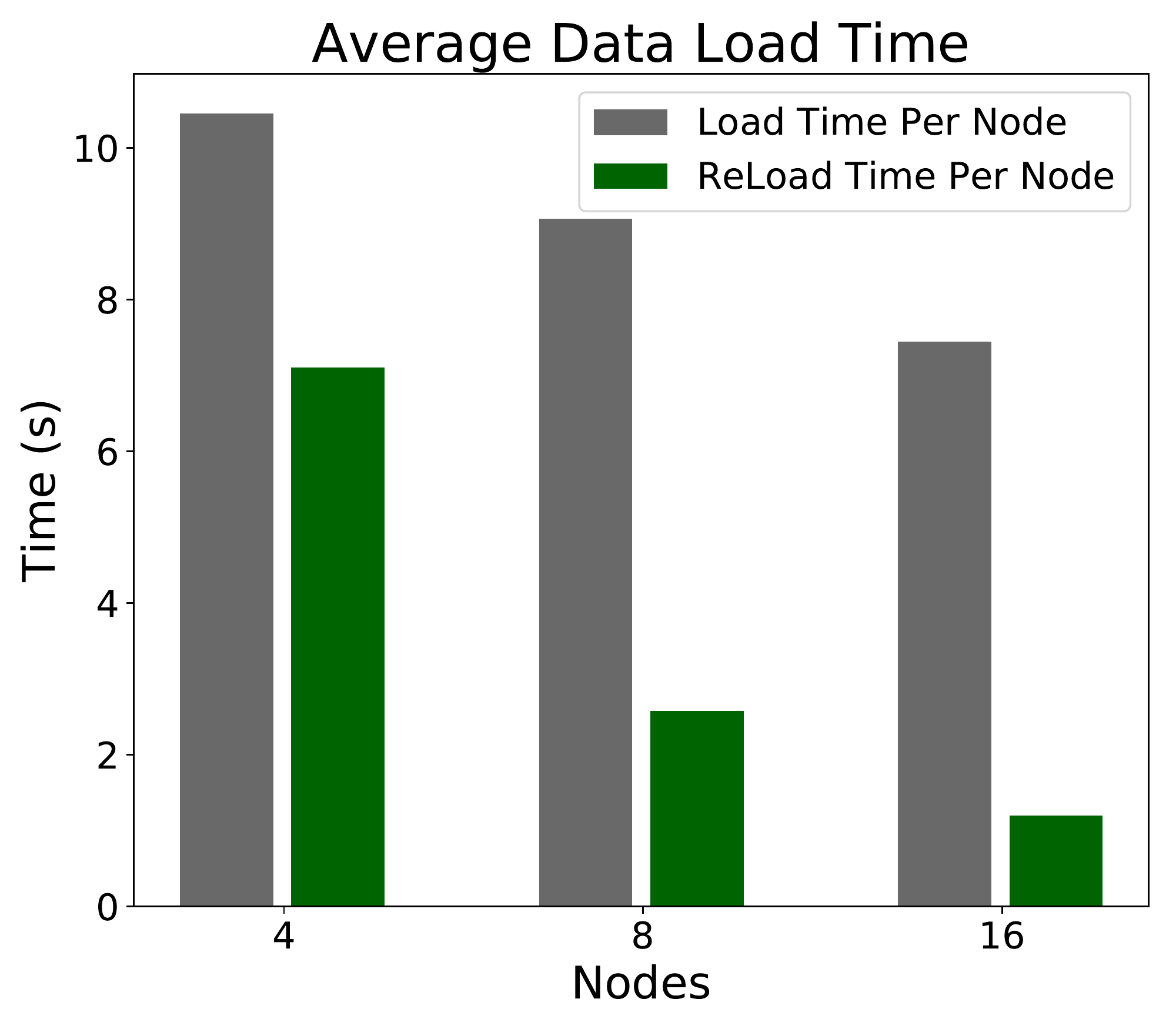}\label{fig:alexnet_dataload_time}}
\subfloat[GoogleNet Data Load Times]{\includegraphics[width=0.25\textwidth]
          {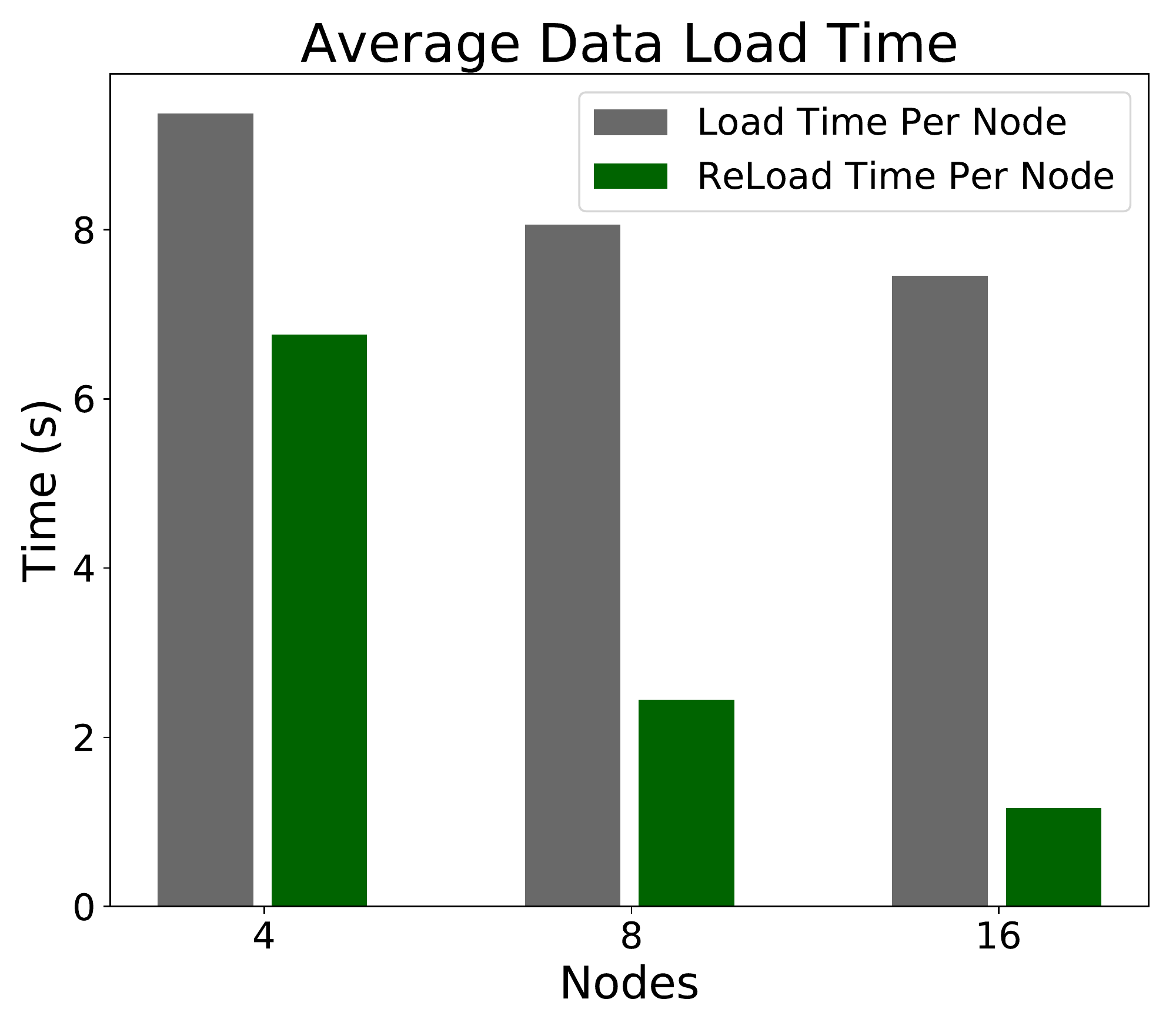}\label{fig:googlenet_dataload_time}}
\subfloat[CIFAR-10 Data Load Times]{\includegraphics[width=0.25\textwidth]
          {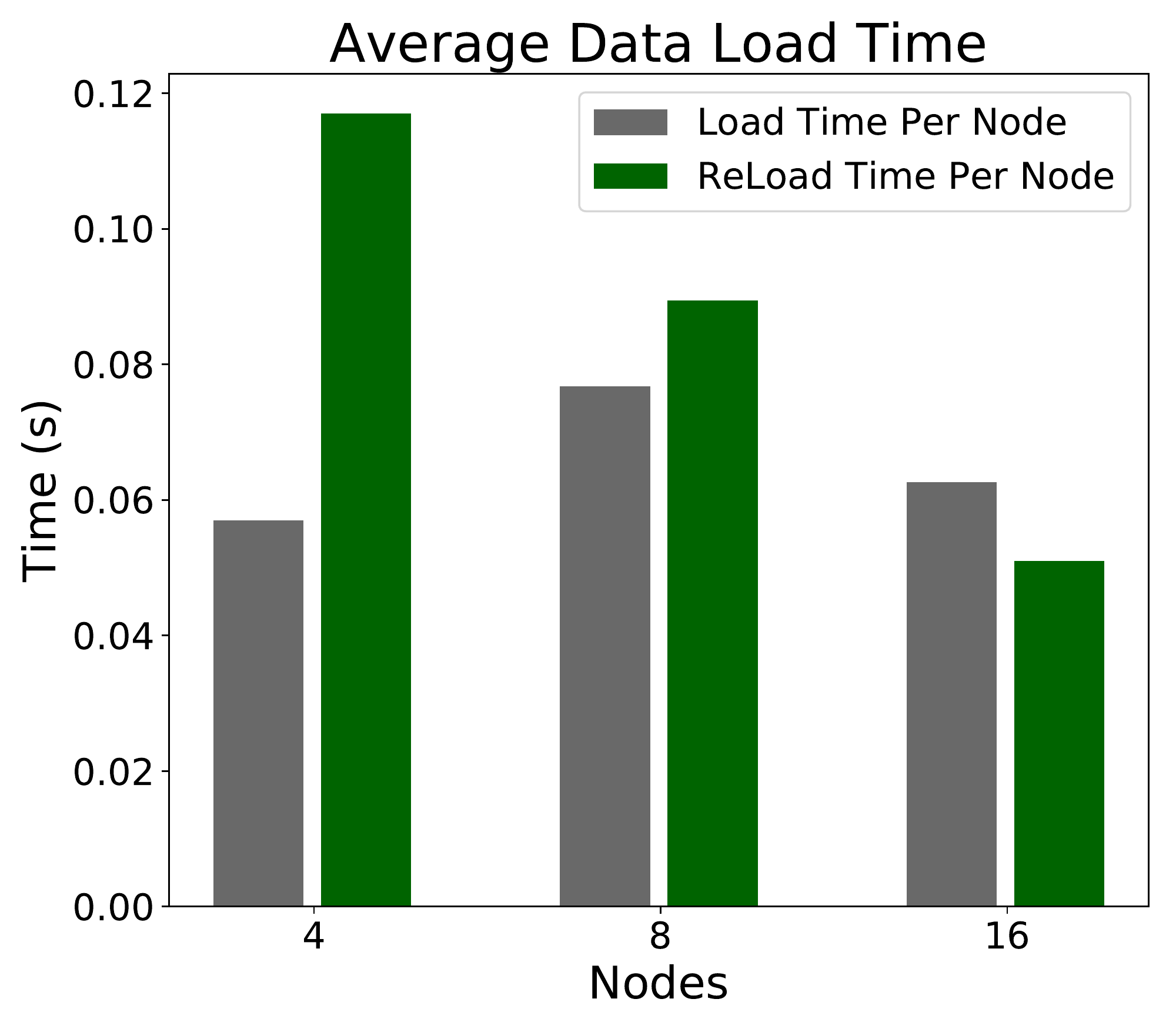}\label{fig:cifar10_dataload_time}}
\subfloat[MNIST Data Load Times]{\includegraphics[width=0.25\textwidth]
          {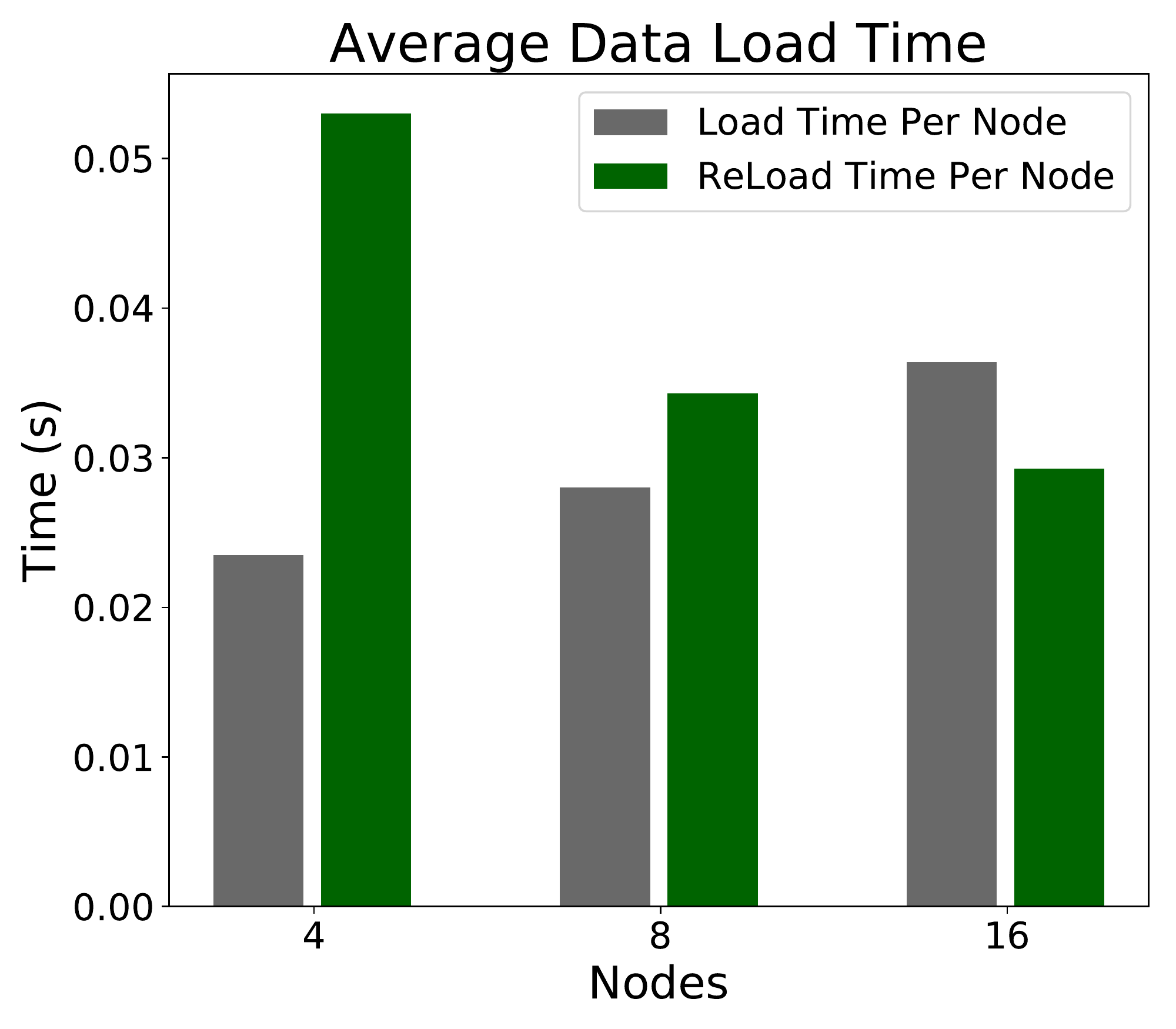}\label{fig:mnist_dataload_time}}
\caption{Time for Loading Datasets}
\end{figure*}

Figures~\ref{fig:alexnet_dataload_time},~\ref{fig:googlenet_dataload_time},
~\ref{fig:cifar10_dataload_time} and ~\ref{fig:mnist_dataload_time} shows the
comparison of reading the overall dataset and partial dataset when a fault
occurs for AlexNet, GoogLeNet, and other datasets. These charts are
specifically useful to understand the cost of reading the dataset as a function
of number of compute nodes. We observe that reading the partial dataset is
significantly faster than reading the entire dataset, which is not surprising.
This is especially validated for ImageNet dataset which is much larger than
MNIST and CIFAR10 datasets. We observe that for MNIST and CIFAR10, partial
reading is actually slower, since these datasets are trivially small.

\begin{figure*}[!htbp]
\centering
\subfloat[AlexNet Shrink Time]{\includegraphics[width=0.25\textwidth]
          {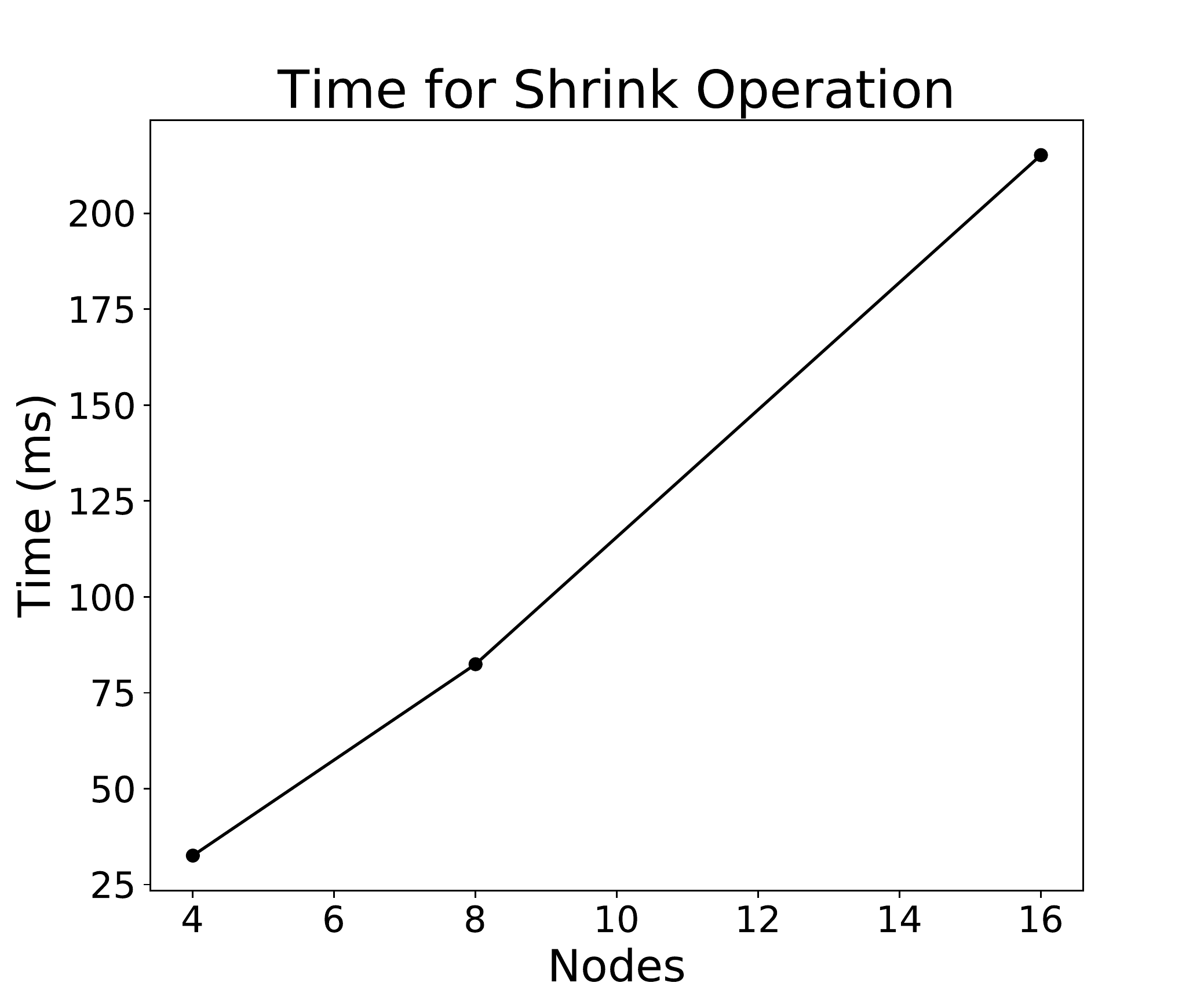}\label{fig:alexnet_shrink_time}}
\subfloat[GoogleNet Shrink Time]{\includegraphics[width=0.25\textwidth]
          {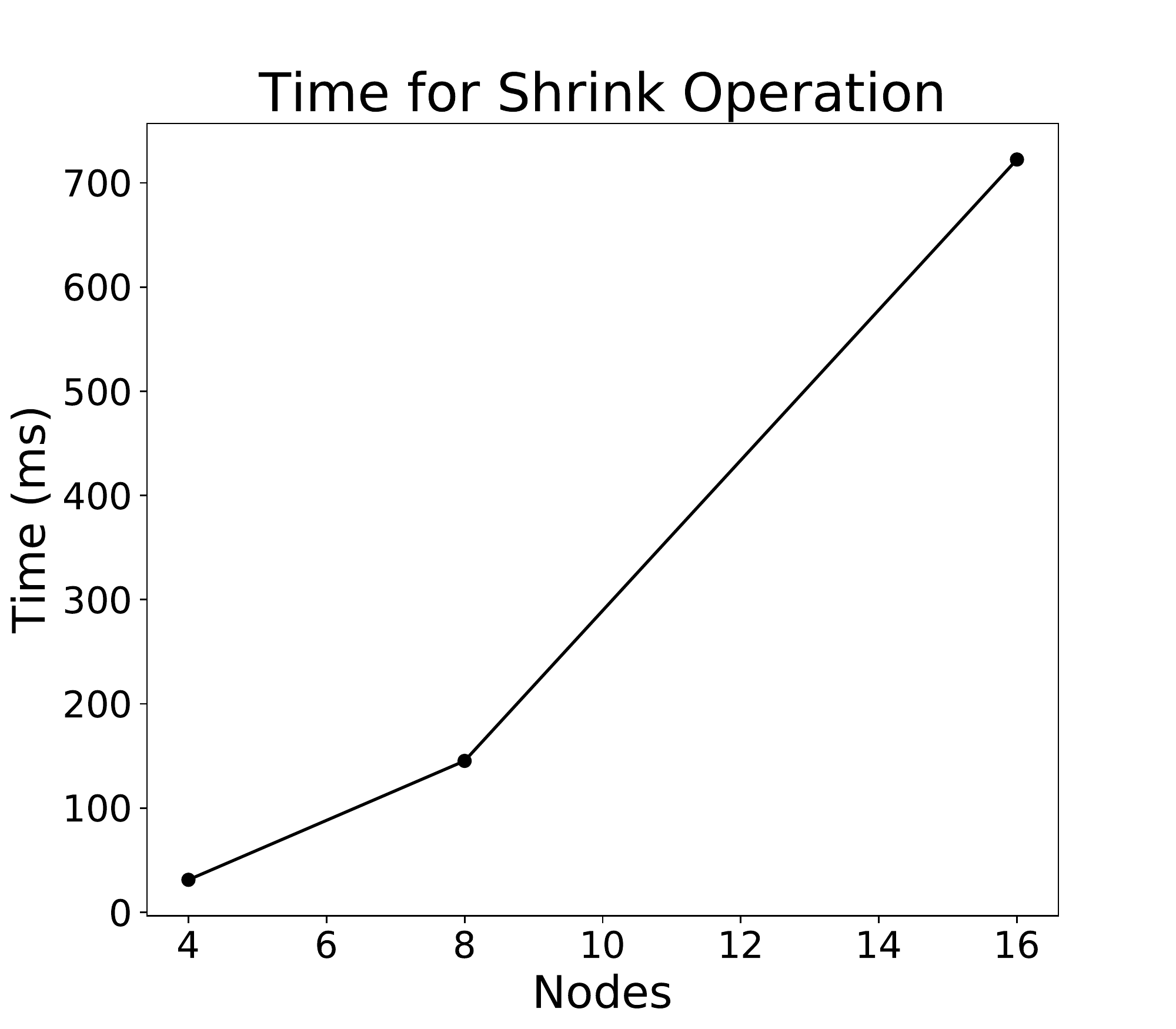}\label{fig:googlenet_shrink_time}}
\subfloat[CIFAR-10 Shrink Time]{\includegraphics[width=0.25\textwidth]
          {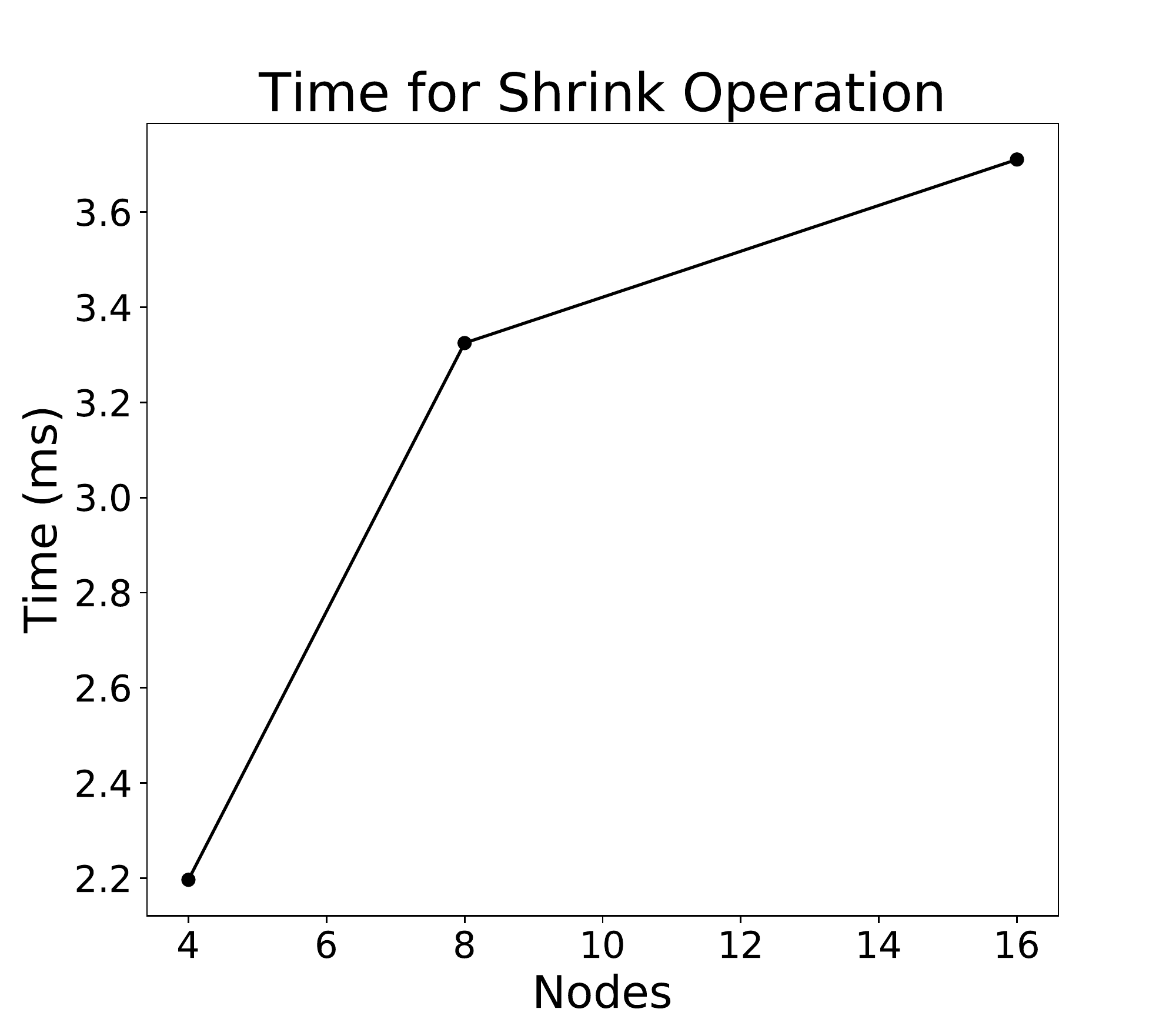}\label{fig:cifar10_shrink_time}}
\subfloat[MNIST Shrink Time]{\includegraphics[width=0.25\textwidth]
          {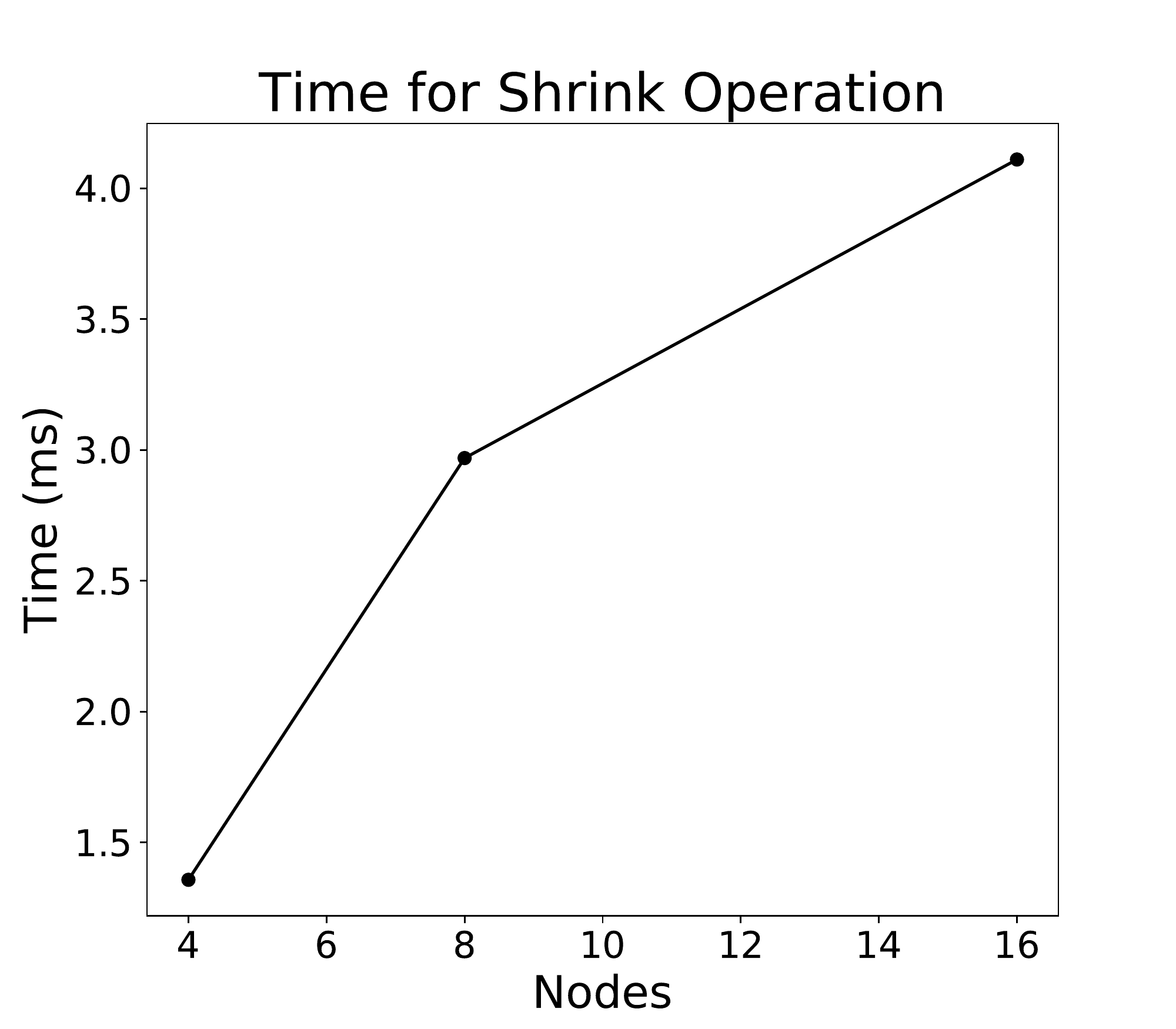}\label{fig:mnist_shrink_time}}
\caption{ULFM Shrink Time}
\end{figure*}

Figures~\ref{fig:alexnet_shrink_time}, ~\ref{fig:googlenet_shrink_time},
~\ref{fig:cifar10_shrink_time} and ~\ref{fig:mnist_shrink_time} shows the
overhead of cumulative \texttt{MPI\_Comm\_shrink} as a function of number of
compute nodes. These functions are executed only if a fault is detected,
otherwise this code is not executed. We observe that the overhead increases
with the number of compute nodes, which is expected. However, the overall time
is relatively insignificant to the batch update time. Hence, the ULFM
specification and the ULFM implementation are sufficient for providing
functionality and performance in implementing DL algorithms.

\section{Related Work}
\label{sec:related}
\label{checkpoint_restart}

The majority of fault tolerant solutions proposed in the literature and
practice have focused on checkpoint-restart mechanisms.
Under these solutions, the applications periodically
save the state of the data and computation either explicitly or implicitly by
using OS level approaches such as Berkeley Lab Checkpoint Restart (BLCR) or
virtualization based approaches.  As an example, ~\cite{Hursey:4228333, Ma:5289172},
primarily focus on checkpoint-restart method for fault
tolerance, storing the checkpointed data into the filesystem, while
~\cite{Gamell:InMemory:7576485}
have presented disk-less in-memory checkpointing storage-restart scheme
at application level leveraging ULFM. Others ~\cite{Ouyang:CRFS:6047205}
have focused on alleviating filesystem I/O bottleneck due to checkpointing,
using other libraries for checkpointing; ~\cite{Kutlu:6507503} explores an
algorithm based fault tolerance for data intensive algorithms with data
replication techniques. Others
~\cite{Walters:4633353} have explored asynchronous decentralized replication with
standard checkpoint restart techniques.  Wang {\em et
al.}~\cite{Wang:Hybrid:5695644} discusses hybrid checkpointing, alternating
between incremental checkpointing and full checkpointing, resulting
in minimized checkpointed data size.
All these techniques have checkpoint-restart as the fundamental
method for providing fault tolerance.
Guo {\em et al.}~\cite{Guo:SC2015} have discussed detect-resume model for
MapReduce, using MPI, in addition to Reinit model. With
Detect-Resume model, the workload from the faulted process is redistributed to
the remaining nodes. In their approach, the lost work is recomputed from scratch
in the remaining processes, leading to a longer recovery time.
However in our approach, for DL algorithms, there is
no need to recompute the work from the lost process(es), hence recovery time is
greatly reduced.
Chakravorty {\em et al.} in their work~\cite{Chakravorty:IHPC06} have explored the
concept of predictive fault tolerance.
While this approach introduces
lesser overhead compared to the checkpoint-restart, the suitability of predictive approach is limited.

\section{Conclusions}
\label{sec:conclusions}
In this paper, we have addressed the question of the requirements of MPI for
designing fault tolerant Deep Learning (DL) algorithms.  We have presented the
case for several types of parallelism as motivated from common use-cases, and
DNN topologies. We have used the discussion to derive the suitability of fault
tolerance proposals in MPI. We have considered several design choices for
implementing fault tolerant DL implementations. Specifically, we have
considered checkpoint-restart, Reinit (when a fault occurs, re-initialize the
MPI automatically) and user-level fault mitigation (ULFM).  We have implemented
our design using MaTEx-Caffe and leveraged ULFM implementation available from
OpenMPI. We have provided an evaluation of fault tolerant MaTEx-Caffe using
ImageNet-1K dataset and widely studied neural network topologies and datasets
such as AlexNet, GoogLeNet on ImageNet dataset, and MNIST and CIFAR-10 datasets
as well.  Our evaluation has indicated the effectiveness of ULFM  both in terms
of its suitability as a specification and readiness for practical deployments.

\bibliographystyle{IEEEtran}
\balance
\bibliography{bibTexts}
\end{document}